\documentclass[a4paper,fleqn,usenatbib]{mnras}

\usepackage[T1]{fontenc}
\usepackage{ae,aecompl}
\usepackage{graphicx}	% Including figure files
\usepackage{multirow}
\usepackage{txfonts}
\usepackage{float}
\def\aj{AJ}
\def\apj{ApJ}
\def\aap{A\&A}
\def\apjl{ApJL}
\def\apjs{ApJS}

\def\mnras{MNRAS}
\def\araa{ARA\&A}

\def\nat{Nature}
\def\pasp{PASP}

\newcommand{\beacon}{{\sc Beacon}}

\newcommand{\vlos}{$v_{los}$}

\title[The structure of And II]{The structure of Andromeda II dwarf spheroidal galaxy}

\author[Andr\'es del Pino et al.]{Andr\'es del Pino$^{1}$\thanks{E-mail: \mbox{adpm@camk.edu.pl}},
Ewa L. {\L}okas$^{1}$, Sebastian L. Hidalgo$^{2,3}$ and Sylvain Fouquet$^{1}$\\
$^1$Nicolaus Copernicus Astronomical Center, Polish Academy of Sciences, Bartycka 18, 00-716 Warsaw, Poland\\
$^2$Instituto de Astrof\'\i sica de Canarias, Calle V\'\i a L\'actea s/n,
E-38200 La Laguna, Tenerife, Spain\\
$^3$Departamento de Astrof\'\i sica, Universidad de La Laguna, Avda. Astrof\'isico Fco. S\'anchez s/n,
E-38206 La Laguna, Tenerife, Spain}

\begin{document}
\label{firstpage}
\pagerange{\pageref{firstpage}--\pageref{lastpage}}
\maketitle

\begin{abstract}
We analyze in detail the spatial distribution and kinematic properties of two different stellar populations in
Andromeda II (And II) dwarf spheroidal galaxy. We obtained their detailed surface density maps, together with
their radial density profiles. The two populations differ not only in age and metallicity, but also in their
spatial distribution and kinematics. Old stars ($\gtrsim 11$ Gyr) follow a round distribution well fitted by truncated
density profiles. These stars rotate around the projected optical major axis of the galaxy with line-of-sight velocities
\vlos$(r_h) = 16 \pm 3$ km s$^{-1}$ and a velocity gradient of $2.06 \pm 0.21$ km s$^{-1}$
arcmin$^{-1}$. Intermediate-age stars ($\lesssim 9$ Gyr) concentrate in the centre of the galaxy and form an
elongated structure extending along the projected optical major axis. This
structure appears to rotate with a velocity gradient of $2.24 \pm 0.22$ km s$^{-1}$ arcmin$^{-1}$, and around the
optical minor axis. The centres of rotation and kinetic position angles (PA$_{\rm kin}$) of both populations differ.
For intermediate-age stars we obtained PA$_{\rm kin} = 18^\circ \pm 2^\circ$ and for the old ones PA$_{\rm kin} =
63^\circ \pm 3^\circ$ in good agreement with photometric PA measured from isopleths fitted to the photometry. We
conclude that the two stellar populations may not be fully relaxed and thus
be the result of a merger occurred at redshift $z\sim 1.75$.
\end{abstract}


%%%%%%%%%%%%%%%%%%%%%%%%%%%%%%%%%%%%%%%%%%%%%%%%%%

%%%%%%%%%%%%%%%%% BODY OF PAPER %%%%%%%%%%%%%%%%%%

\section{Introduction}\label{introduccion}

The Local Group (LG) of galaxies is a unique laboratory for studying galaxy formation and evolution. Due to their
proximity, the evolution of its member galaxies can be determined with unparalleled accuracy from their formation at
high redshift to the present time. The dominant type among the dwarf galaxies of the LG are dwarf spheroidal (dSph)
systems characterized by low surface luminosities ($\rm \Sigma_V \la 0.002~L_\odot~pc^{-2}$), small sizes (about a few
hundred parsecs) and the lack of gas. The relatively large velocity dispersions observed in dSphs, exceeding 7 $\rm
km~s^{-1}$ \citep[see][and references therein]{Aaronson1983, Mateo1998, McConnachie2012}, suggest the presence of
large amounts of dark matter in them. Assuming that they are in equilibrium, dSphs are found to possess mass-to-light
ratios ($M/L$) of $\sim 5-500$ in solar units (\citealt{Kleyna2001, Odenkirchen2001, Kleyna2005}; \citealt*{Lokas2005}; \citealt{Lokas2009}).
These discoveries have given rise to competing interpretations concerning the origin of these properties and their
cosmological significance.

In the $\Lambda$CDM scenario, dwarf galaxies are the building blocks from which larger galaxies are formed
(\citealt{Blumenthal1985, Dekel1986}; \citealt*{Navarro1995}; \citealt{Moore1998}). The dwarfs we observe today may be surviving systems that
have not yet merged with larger objects and may contain the fossil record of the early evolution of galaxies. The
evolution and star formation history (SFH) of these galaxies are very likely affected by local processes such as
supernovae feedback and tidal interactions with nearby systems, as well as by global cosmic environmental factors like
the early reionization of the Universe by UV radiation (\citealt{Taffoni2003, Hayashi2003}; \citealt*{Kravtsov2004}; \citealt{Kazantzidis2011}).
Some of the dSph galaxies of the LG show great complexity in their SFH and kinematics (see for
example \citealt*{Stetson1998}; \citealt{DaCosta1996, Battaglia2008b, Fraternali2009, deBoer2012, Weisz2014, delPino2017}), which has been associated
with the presence of multiple stellar populations. This is the case of And II, one of the satellites of M31.

Located at the distance of 184 kpc from M31 \citep{McConnachie2012}, Andromeda II (And II) is one of the most luminous M31 dSph
companions. Its colour magnitude diagram (CMD) displays a conspicuous double red giant branch (RGB) which has been
interpreted as the presence of two stellar populations (\citealt{DaCosta2000}; \citealt*{McConnachie2007}). This was later confirmed
by its SFH \citep[][Hidalgo et al. in preparation]{Weisz2014, Gallart2015}, which shows two clearly distinct star
formation bursts that occurred $\sim 12.5$ and $\sim 7-8$ Gyr ago, respectively.

Using the wide field of view of the Subaru Suprime-Cam camera, \citet{McConnachie2007} studied the spatial
distribution of the And II stars by selecting them according to their position in the RGB and the horizontal branch
(HB) of the CMD. They showed that the bluer population was more spatially extended than the redder one, concluding that
And II is indeed composed of two different structural components. Furthermore, And II shows a strong rotation signal,
comparable to its central velocity dispersion. This feature, not common in dSph galaxies, is even more surprising
since the rotation direction is around the optical major axis of the galaxy \citep{Ho2012}. This kinematic signature was
later confirmed by \citet*{Amorisco2014}, who also claimed to have found a stream of stars located at $\sim$ 1.5 kpc
from the centre.

Some efforts have been undertaken to explain the origin of these unusual properties. \citet{Lokas2014b}
proposed an evolutionary model in which And II forms as a result of a head-on merger between two similar late-type dwarf
galaxies. In this scenario the different spatial distributions of the two populations are explained as due to slightly
different sizes of the progenitor disks. In a follow-up study, \citet{Ebrova2015} demonstrated that prolate rotation can
originate from a variety of orbital initial conditions of the merger but would be impossible to produce via tidal
stirring of a disky dwarf orbiting a larger host. Recently, \citet{Fouquet2017} extended the model to include gas
dynamics and star formation which allows to reproduce not only the kinematics but also the SFH and the spatial
distribution of different populations. They have also modelled the effect of ram pressure stripping in the hot halo of
M31 which leaves And II devoid of gas in agreement with observations.

In the present paper we reanalyze the Suprime-Cam camera data for And II, but with the benefit provided by the detailed
SFH of the galaxy (Hidalgo et al. in preparation), and the spectroscopy of $\sim550$ RGB stars from
\citet{Ho2012}. We use the detailed SFH of the galaxy to assign ages and metallicities to the stars present in the
\citet{McConnachie2007} photometry. This allows us to obtain precise surface density maps of the stellar content of And
II, and to study the spatial distribution of both stellar populations in unprecedented detail. These maps are
compared with the velocity maps obtained from the spectroscopy in order to derive their kinematic properties.

The paper is organized as follows. In section~\ref{Cap:Data} we present the data set used in this study.
Section~\ref{Cap:Methodology} describes the method used to derive stellar properties using the CMD.
Section~\ref{Cap:CMD} presents the CMD of And II and its features. In section~\ref{Cap:2DMaps}, we discuss the
2-dimensional distribution of the stellar populations. Section~\ref{Cap:Radial_Profiles} explains the
procedure for obtaining the radial density profiles and the results from isopleth and model fitting are
discussed. In section~\ref{Cap:Spectra}, we present the results from spectroscopy and compare them to those from
the photometry. In section~\ref{Cap:Discussion}, we discuss the results and search for signatures of
possible interactions with other systems in the data. Lastly, a short summary and the main conclusions of this work are
presented in section~\ref{Cap:Summary}.

\section{The data}\label{Cap:Data}

And II galaxy has been thoroughly observed over the last decade within different projects and with different aims. Many
of these observations have produced high quality data which have helped to improve our knowledge about the galaxy. In
the present work, we focus on the study of its stellar content through photometry and spectroscopy of its resolved
stars. For the purpose of this study, deep photometry and a wide field of view are the two most desired features.
Deep photometry, reaching the oldest main sequence turn-offs (oMSTO) at high completeness ($> 75\%$), is required in
order to derive the ages and metallicities of the stars i.e. the SFH of the galaxy. Wide field-of-view photometry
(WFOVP) is necessary to properly study spatial features and gradients present among different stellar populations. In
the case of And II there is no photometric catalog fulfilling both requirements, encouraging us to combine information
from several previous studies.

\begin{figure}
\begin{center}
\includegraphics[scale=0.7]{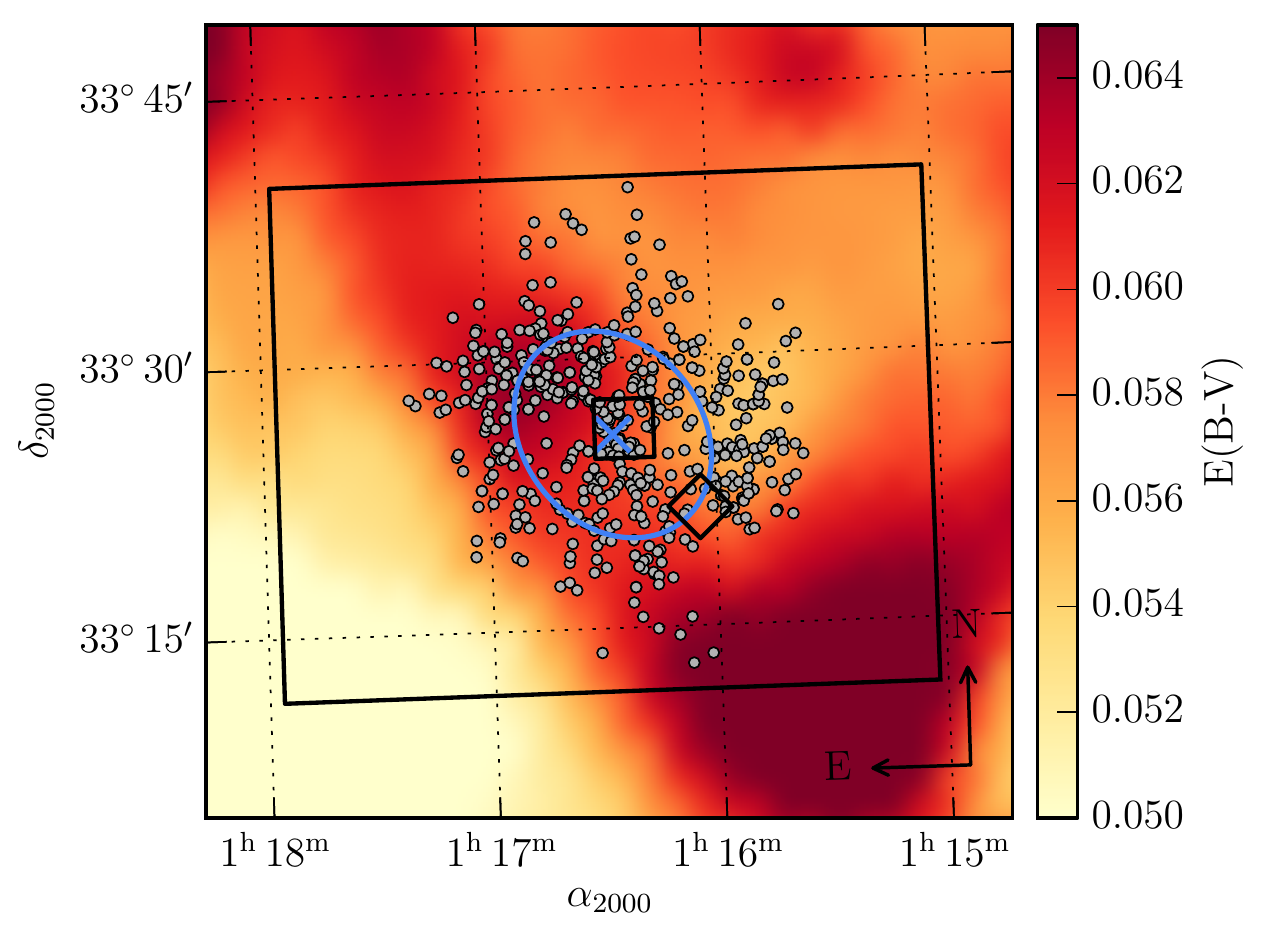}
\caption[A schematic view of the wide field photometry data]{A schematic view of the spatial coverage of the data sets.
The large rectangle shows the area covered by the WFOVP from \citet{McConnachie2007}. Smaller black squares mark the
position of the centre-most ACS@HST field and the outer WFC3@HST field, {observed in parallel mode} \citep{Weisz2014}.
Small grey circles indicate the position of the member stars for which there is spectroscopy available \citet{Ho2012}.
The blue ellipse represents the half-light radius of And II ($6.2^\prime$) with a position angle of 37$^{\circ}$ and an ellipticity
$\epsilon = 0.2$ (see Table~\ref{tab:And_II_final}). The centre of And II is marked by a cross. A reddening map of
the observed region is shown in the background \citep{Schlafly2011}.}
\label{fig:Data}
\end{center}
\end{figure}

We have combined three different kinds of data kindly provided by the authors mentioned below: the detailed SFH derived
from deep (F475W $\sim 29$) HST photometry \citep[][Hidalgo S. L. private communication]{Weisz2014}, the Subaru  WFOVP
obtained by \citet{McConnachie2007}, and lastly the \citet{Ho2012} catalogue of {$531$ members RGB stars} for which
line-of-sight velocities (\vlos) and metallicities ([Fe/H]) were derived. In Figure~\ref{fig:Data} we show the spatial
coverage of the collected data.

We corrected the Subaru WFOVP list for distance modulus and reddening. A distance modulus of $(m-M)_0=24.07 \pm 0.06$
($652 \pm 19$ kpc) was adopted for And II using the median of all distance measurements in the literature provided by
the NED database. The reddening correction was applied individually over each star by cross-correlating their positions
with the dust maps of \citet*{Schlafly2011} {(see Figure~\ref{fig:Data})}. Finally, we cleaned our photometry of likely non-stellar objects and stars
with high magnitude errors using their photometric errors ($\sigma$) and quality flags based on the deviation of the stellar locus.
{We used the membership selection performed by \citet{Ho2012} on the spectroscopic sample. This is based on the position of the stars in the CMD, their \vlos, and the strength of the Na I absorption line at $\lambda8190 \AA{}$.}

\section{Obtaining ages and metallicities from the CMD}\label{Cap:Methodology}
\subsection{The averaged SFH}\label{Cap:Methodology:What_we_have:The_averaged_SFH}

The CMD of a galaxy provides valuable information about different stellar populations present in the stellar system.
The positions of the stars in the CMD depend on both their intrinsic characteristics (i.e. mass and metallicity)
and their evolutionary state. This also means that in the case of a complex stellar population the CMD will suffer from
the age-metallicity degeneracy. This degeneracy must be broken in order to be able to assign ages and
metallicities to the stars based on their position in the CMD. To do so, and derive the ages and metallicities for the
WFOVP stars, we have used the information provided by the SFH of the galaxy.

Following the procedure explained in \citet*{delPino2015}, both data sets were combined in order to resolve
the age and metallicity from the deep photometry and to get spatial coverage from the WFOVP. A comprehensive
description of the method is given in the aforementioned paper, while here we give only a short summary.

Only two small fields (ACS and WFC3) are available with sufficient depth to derive the SFHs. One of them is
located at the centre of And II, the other one approximately at $\sim6.65^\prime$ in the south-west
direction from the former, just outside the half-light radius of the galaxy, $r_{h} = 6.2 \pm2 ^\prime$
\citep{McConnachie2012}. Despite the relatively small area covered by the deep photometry ($\sim2\%$ of the
area covered by the WFOVP), the two fields span enough galactocentric radius ($7.5^\prime$) to build a reliable
axisymmetric spatial model of the SFH for the galaxy. While this model will not reproduce any local spatial feature in
the SFH, it gives a good approximation of the expected average SFH for the whole galaxy.

\begin{figure}
\begin{center}
\includegraphics[trim=0.cm -0.22cm 0.cm 0.cm, clip=true, scale=0.7]{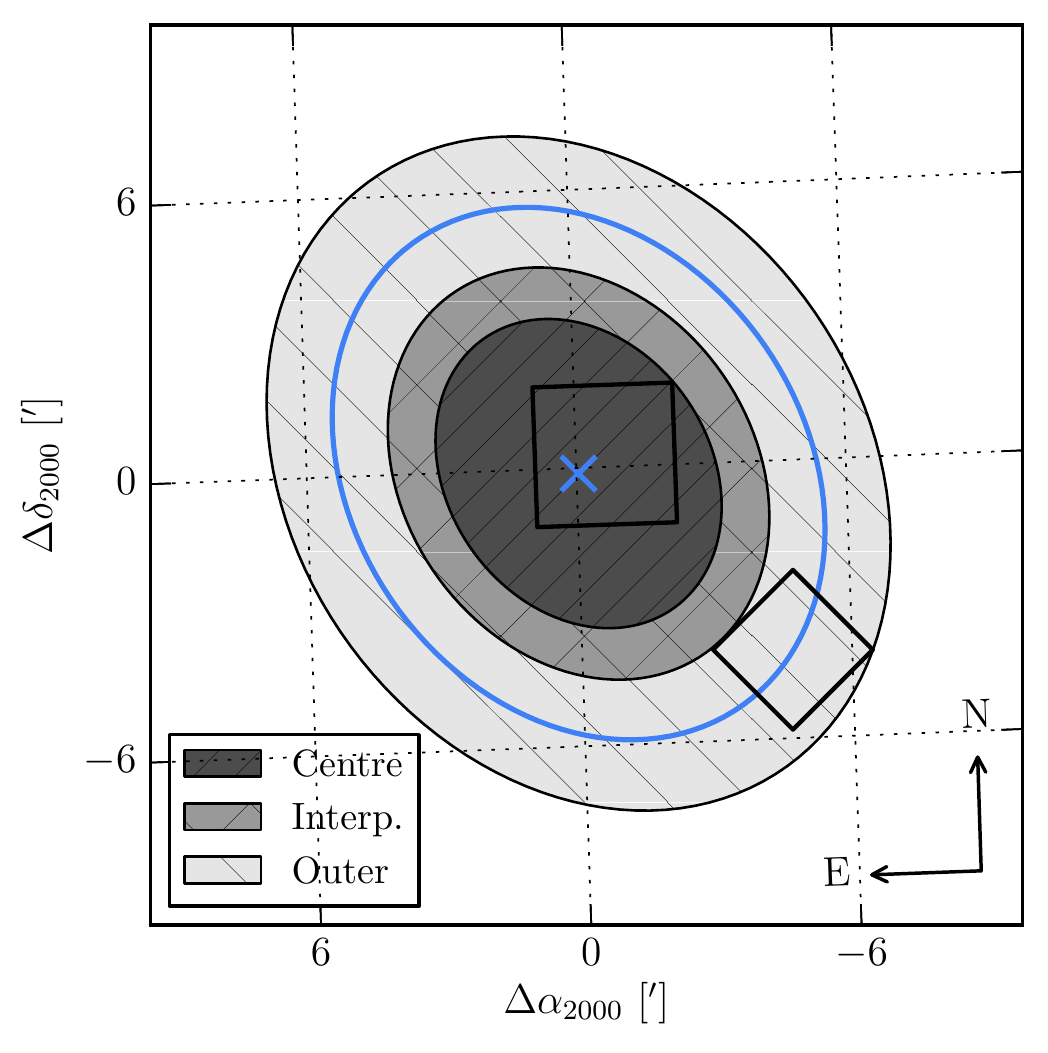}
\caption[Adopted axisymmetric model for deriving the global SFH]{Schematic view of the axisymmetric radial model
adopted for estimating the global SFH of And II. The fields in which the SFH was derived are represented by black
squares. Different shaded annular regions show the regions in which the SFHs of the centre-most field and the outer
one were extrapolated. Symbols and line types are the same as those used in Figure~\ref{fig:Data}.}
\label{fig:Averaged_Radial_model}
\end{center}
\end{figure}

\begin{figure}
\begin{center}
\includegraphics[scale=0.7]{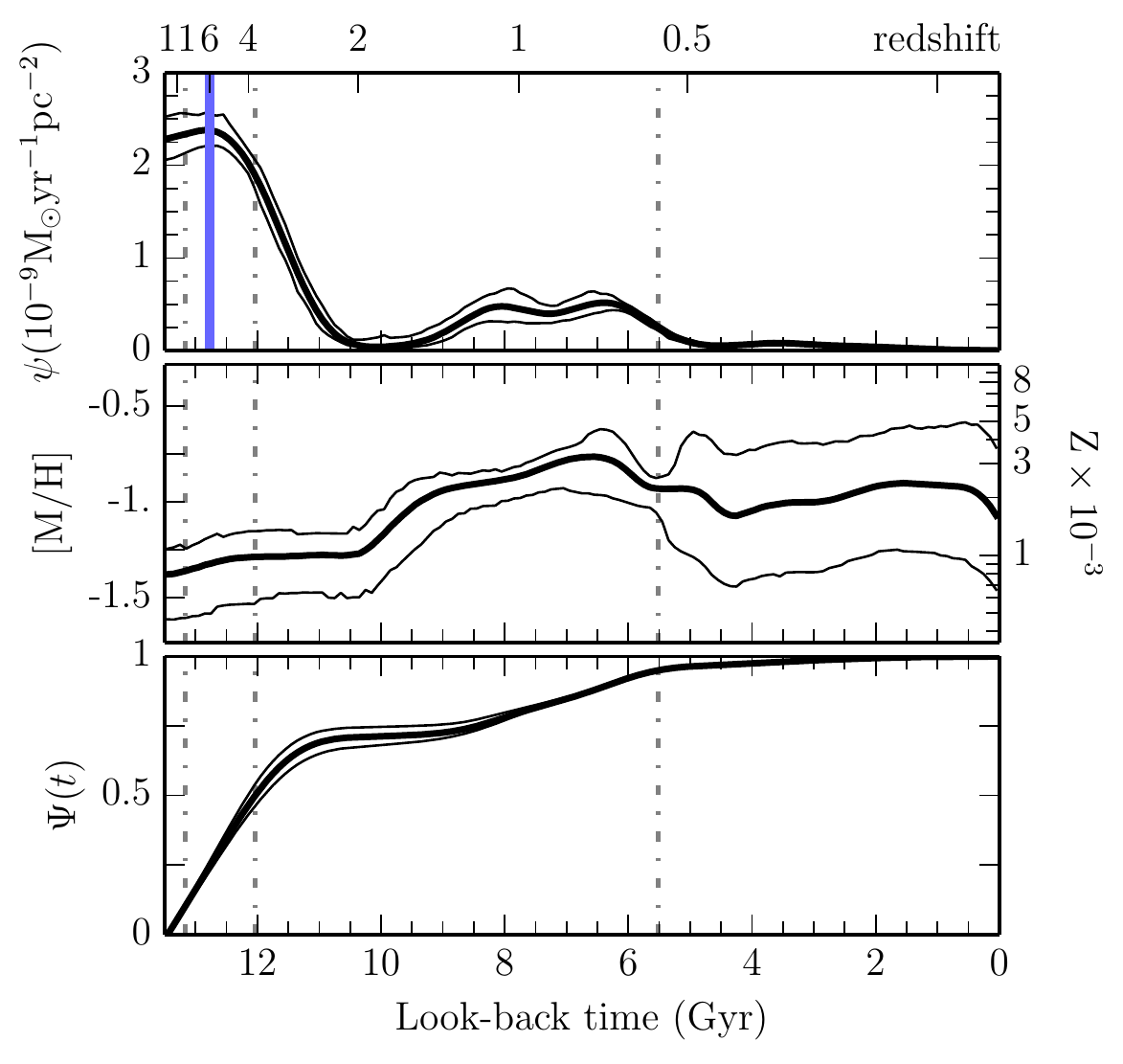}
\caption[The SFR, AMR, and cumulative mass fraction as a function of time for the global SFH]{The star formation rate
as a function of time ($\psi(t)$) (top), the AMR (middle), and the cumulative mass fraction, $\Psi(t)$, (bottom) of the
radial model of And II (Figure~\ref{fig:Averaged_Radial_model}). Error bands in $\psi(t)$ and the dispersion in the AMR
are drawn by thin lines. Units of $\psi(t)$ are normalized to the total area covered by the model. The 10th, 50th, and
95th percentiles of $\Psi(t)$ are shown by dash-dotted vertical lines. The UV reionization era is marked at $z\sim6$
by a blue vertical line. The redshift scale given in the upper horizontal axis was computed assuming $H_0=70.5$ km
s$^{-1}$Mpc$^{-1}$, $\Omega_{\rm M}$=0.273, and a flat universe with $\Omega_\Lambda$=$1-\Omega_{\rm M}$.}
\label{fig:Andromeda_II_model}
\end{center}
\end{figure}

We constructed the model assuming that the SFH of the galaxy is elliptically symmetric. We define two concentric
elliptical regions, namely Centre and Outer, centred at the galaxy with ellipticity $\epsilon = 0.2$
\citep{McConnachie2012} and covering the two observed HST fields. The SFH is assumed to be constant within each of
these regions, and equivalent to the one derived in its corresponding observed field. The SFH of the annular region
between Centre and Outer (Interpolation) was calculated as the linear interpolation between {the SFH obtained in the} two probed regions.
Finally, the total average SFH of the whole galaxy was computed as the arithmetic mean of the three SFHs derived within
the three concentric elliptical regions, weighted by their areas. Figure~\ref{fig:Averaged_Radial_model} shows the
axisymmetric model used to compute the averaged SFH. Some relevant integrated and averaged quantities associated with
it are listed in Table~\ref{tab:Integrated_Quantities_2}. Figure~\ref{fig:Andromeda_II_model} shows the averaged star
formation rate (SFR), the age-metallicity relation (AMR) and the total cumulative SFH we computed for And II. Two
stellar populations can be clearly seen in the SFH. The first one represents $\sim$ 70\% of the total mass converted
into stars and it consists of old ($\sim 12.5$ Gyr), relatively metal-poor ([Fe/H]$\sim -1.45$) stars. A second burst of
star formation produced an intermediate-age population (9-5 Gyr) of more metal-rich ([Fe/H]$\sim -0.9$) stars.
Interestingly, the second burst of star formation appears to be a blend of two not well resolved bursts, one
occurring $\sim 8$ Gyr ago and followed by another $\sim 6.25$ Gyr ago.

\begin{table}
\begin{minipage}{\columnwidth}
\caption{Integrated quantities derived from the radius-dependent model of SFH in the And II
dSph.}
  \label{tab:Integrated_Quantities_2}
 \centering
  \begin{tabular}{@{}lcccc}
    \hline
    \hline
      Region  & $\int{\psi(t'){\rm d}t'}$ \footnote{Integrated between 0 and 13.5 Gyrs.}& $\langle\psi(t)\rangle$  & $\langle\rm age\rangle$ & $\langle Z \rangle$ \\
      & $[10^7$ M$_\odot]$ & $[10^{-10}$ M$_\odot \rm yr^{-1} pc^{-2}]$ & $\rm[Gyr]$ & $\times10^{-3}$\\
      \hline
      Centre       & $1.3 \pm 0.3$   &  $7 \pm 2$      & $10.6 \pm 1.2$ & $1.6 \pm 0.5$ \\
      Interp.      & $0.8 \pm 0.2$   &  $5 \pm 1$      & $10.8 \pm 1.0$ & $1.4 \pm 0.3$ \\
      Outer        & $1.0 \pm 0.2$   &  $2.3 \pm 0.6$  & $11.2 \pm 0.9$ & $1.2 \pm 0.2$ \\
      Total        & $3.0\pm 0.4$    &  $4 \pm 1$      & $11.0 \pm 1.8$ & $1.3 \pm 0.6$ \\
    \hline
  \end{tabular}
\end{minipage}
\end{table}

\subsection{Sampling the CMD with the sCMD}\label{Cap:Methodology:What_we_have:Sampling_the_CMD}

{We used the averaged SFH as input for the \textsc{IAC-Star} code \citep{Aparicio2004} to compute
a synthetic CMD (sCMD) of $5\times10^7$ stars, corresponding to the averaged SFH of And II. This sCMD is therefore
comparable to the CMD of the WFOVP, but includes some of the intrinsic properties we know from
stellar evolution libraries, such as ages, metallicities or masses of stars. We can use this information to infer the
properties of the observed stars by comparing both CMDs, the observed and the synthetic one.}

Before performing any quantitative comparison between both CMDs, \textit{observational effects} affecting the observed
CMD must be simulated in the sCMD. These include signal-to-noise limitations, stellar crowding, detector defects and
all factors affecting and distorting the observational data with the resulting systematic uncertainties, changes in
measured colours and magnitudes, and the loss of stars. {The most precise way to quantify the observational
effects is through artificial star tests \citep[see][for an example]{Hidalgo2011}. This was the procedure followed
in the HST fields, in order to derive the SFH. However, this process requires large computer resources besides the considerable effort
of re-obtaining the photometry from the original images.

In order to assign ages and metallicities to the stars in the WFOVP, we do not require
such precision in the errors simulation over
the sCMD. Instead, we simulated observational errors shifting the magnitudes of every synthetic star according to the photometric errors
observed in their position in the CMD of the WFOVP. For further information we refer the reader to \citet{delPino2015}. This allows a realistic enough
comparison between the observed and synthetic CMDs.}

\begin{figure}
\begin{center}
\includegraphics[scale=0.7]{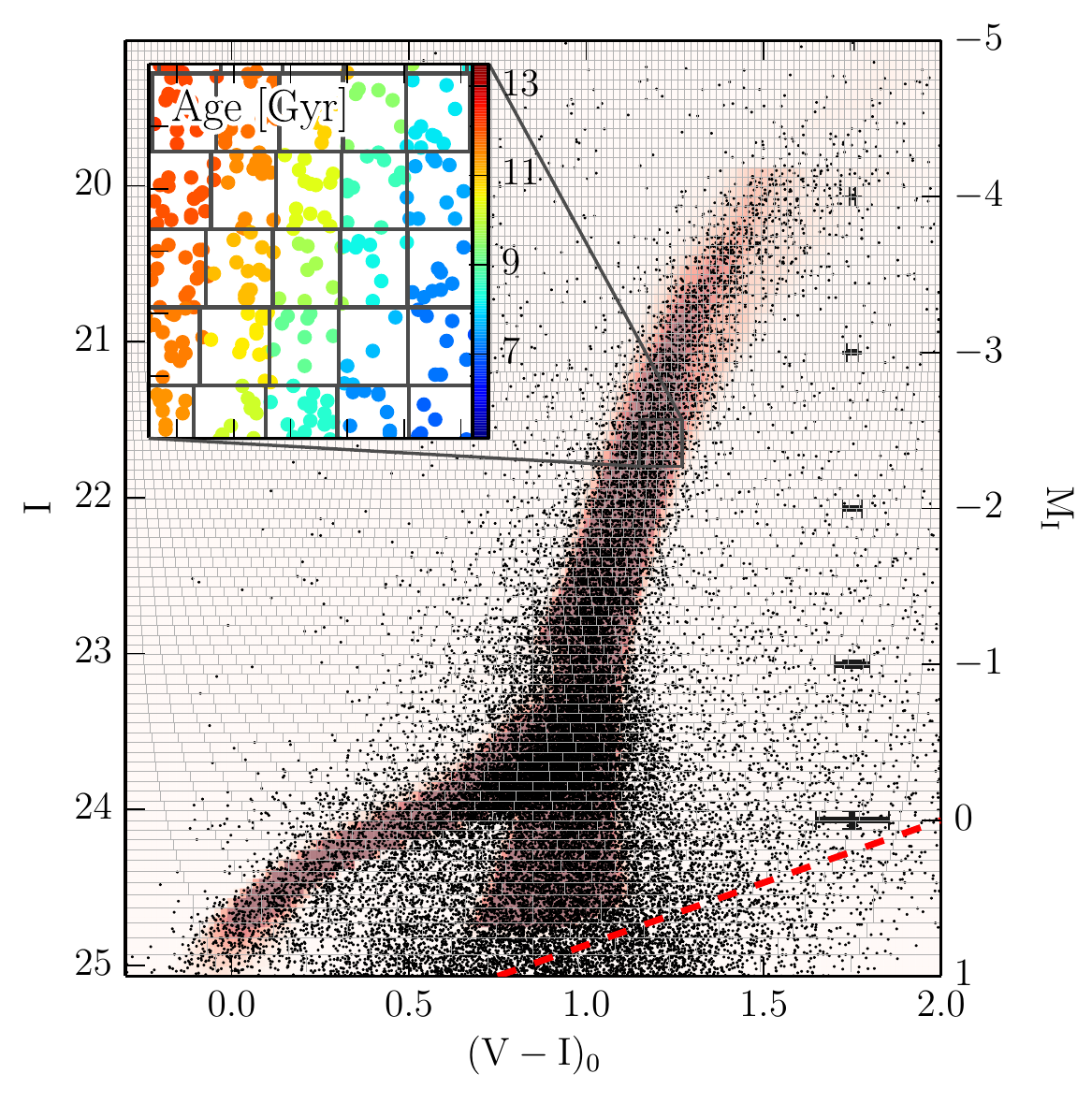}
\caption[]{Sampling performed between the observed CMD (black dots) and the synthetic one with the simulated
observational effects (esCMD) in red. {The CMD was obtained from the photometry derived by \citet{McConnachie2007}, while
the esCMD is the one corresponding to the SFH shown in
Figure~\ref{fig:Andromeda_II_model}}. Photometric errors are represented by the error bars on the right. The limit
adopted for the sampling is indicated by the red dashed line, corresponding to approximately the 50\% of completeness
level in photometry. The zoomed region shows how stellar properties are assigned to the stars inside the sampling
cells. Stars lying within a given cell will share the same age and metallicity, obtained from the esCMD. Here, the
colours represent the assigned ages.}
\label{fig:Grid}
\end{center}
\end{figure}

The observed CMD and the sCMD with the simulated observational effects (esCMD) are then compared by an
adaptive sampling procedure. For this purpose, a sampling grid is generated based on photometric errors from the
observed CMD and on the number of sampled stars per bin. The grid is composed of 7804 cells in total, with sizes
varying from $\Delta I \approx 0.063$ and $\Delta(V-I) \approx 0.018 $ to $\Delta I \approx 0.135$ and $\Delta(V-I)
\approx 0.25 $, chosen in order to encompass the photometric errors. In Figure~\ref{fig:Grid} we show
the grid used for sampling both CMDs. Observed and synthetic stars lying within a specific cell are assumed to have
similar properties. We adopted the average of the relevant properties as the common value for all
observed stars falling within a given bin, while their standard deviations give us an estimate of the error of these
adopted values.

In order to minimize the dependence of the solution on photometric zero points, reddening and the distance to the
galaxy, we sampled both CMDs according to the procedure explained above for 400 different offsets in colour and
magnitude applied to the observed CMD. The offsets span $\Delta(m_v-m_i)=(-0.1, 0.1)$ and $\Delta(I)=(-0.25, 0.25)$
in colour, in steps of $\Delta(m_v-m_i)= 0.0125$ and $\Delta(I)=0.01$ respectively. For each offset, we compared the
obtained results for the synthetic and the observed CMD by counting the stars within the cells used for the sampling,
and the final age-metallicity distributions obtained. The offset providing best results, with its associated
dispersion, was adopted as the final distance modulus and reddening values for And II. The final adopted shifts were
$\Delta(m_v-m_i)= 0.05 \pm 0.02$, $\Delta(I)= -0.15 \pm 0.09$.

\subsection{Errors and consistency test}\label{Cap:Methodology:Consistency}

\begin{figure}
\begin{center}
\includegraphics[scale=0.7]{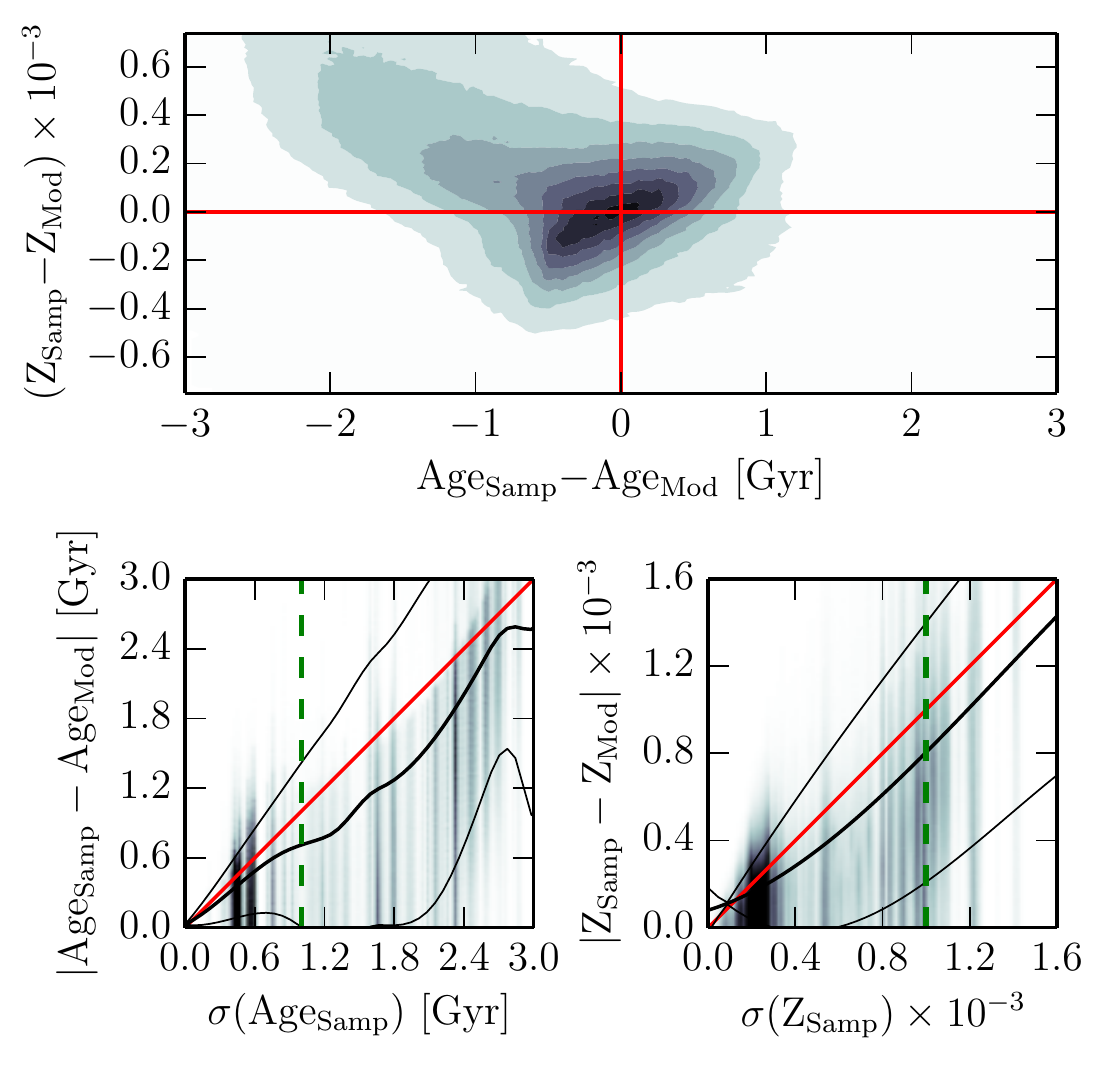}
\caption[]{Errors in the sampled Ages (Age$_{\rm Samp}$) and metallicities (Z$_{\rm Samp}$). Top panel: distribution
of total errors in the sampled age and metallicity. Bottom panels: absolute errors in age and metallicity as a function
of the dispersion of these quantities measured over the CMD. The red line shows a one-to-one relation. Black curves
plot the average of the real error as a function of the sampled one. Thin lines show the $1\sigma$ dispersion of these
values. Vertical green dashed lines mark the maximum adopted errors for the selection of the stellar
populations (section~\ref{SubCap:2DMaps:Desity_distribution}).}
\label{fig:Errors}
\end{center}
\end{figure}

\begin{figure*}
\begin{center}
\includegraphics[scale=0.7]{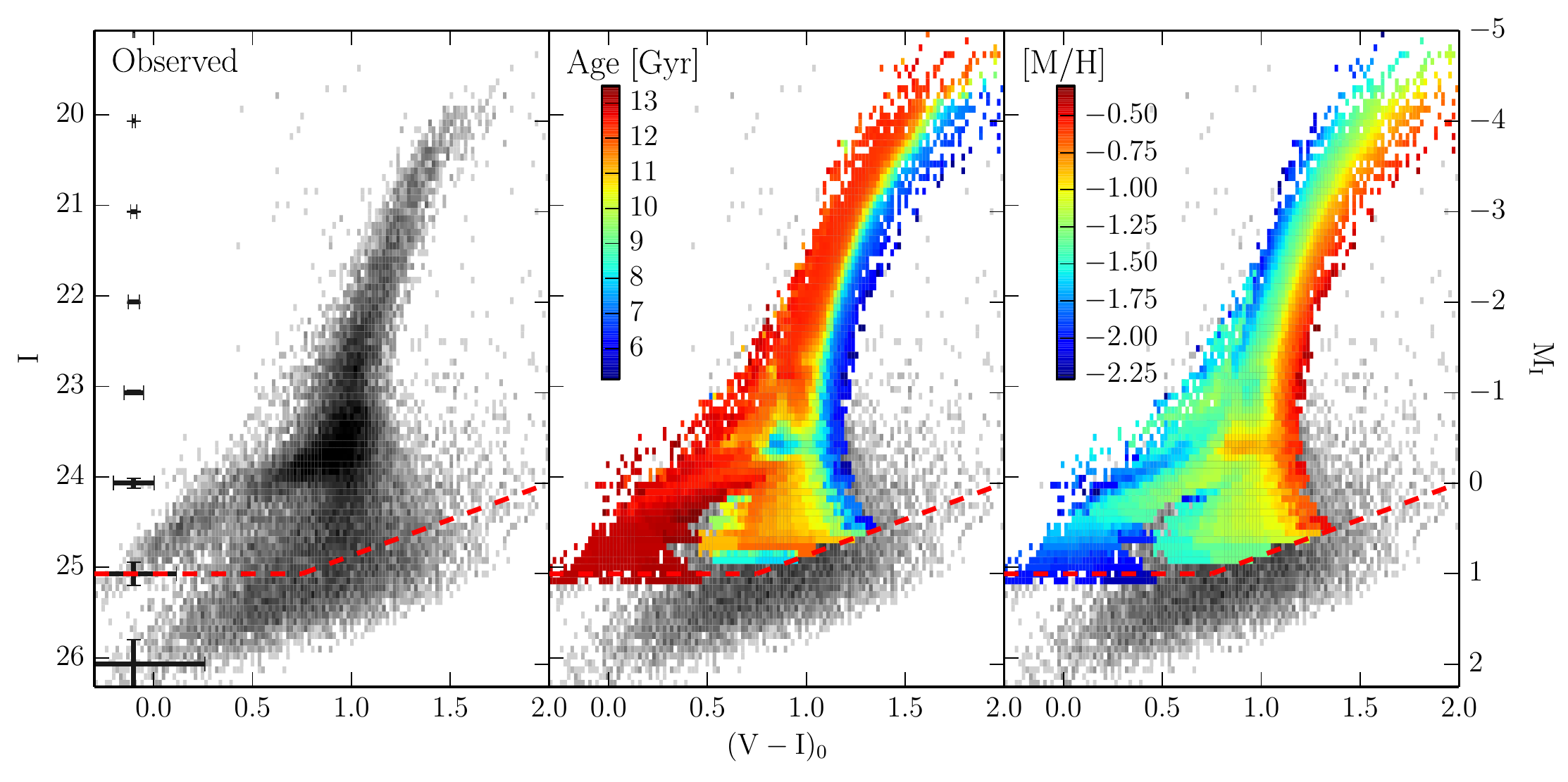}
\caption[CMD of And II]{Left panel: Hess diagram of the calibrated CMD of And II, obtained from the photometry derived by \citet{McConnachie2007}.
The shading indicates the concentration of stars in logarithmic scale. Error bars on the left show the size of the observational errors. Middle
panel: recovered ages in Gyr after the sampling. The colour scale indicates age, while grey areas represent stars which
did not fulfil our completeness requirements during the sampling. Right panel: same as middle panel, but for the
metallicity ([Fe/H]). In all panels the red dashed line marks the boundary imposed on our sampling, equivalent to the
completeness level of 50\%.}
\label{fig:CMD}
\end{center}
\end{figure*}

Stars were tagged in the computed sCMD in order to follow them through the whole procedure of error simulations and
sampling. This allows us to study how errors propagate and affect our age and metallicity determination. These errors
include uncertainties in the photometric zero points, reddening, distance to the galaxy, errors in the magnitudes of
the stars, and uncertainties in the SFH. To analyze their influence, we created $10^3$ esCMDs based on different SFHs,
each one with SFRs and AMRs randomly selected from a normal distribution centred on the original SFR and AMR and
$\sigma$s equal to their errors. In addition, small random shifts in colour and magnitude were added according to the
dispersion of the best offset found in the previous section for the CMD. Lastly, the $10^3$ esCMDs were sampled using the
original esCMD as a reference model. The sampled ages (Age$_{\rm  Samp}$) and metallicities (Z$_{\rm  Samp}$) obtained
for the esCMD were then compared with the original ages (Age$_{\rm  Mod}$) and metallicities (Z$_{\rm Mod}$) of the
stars from the sCMD.

The resulting average differences were $\langle \rm Age_{Samp} - Age_{Mod}\rangle$ = $-7.2\times10^{-6}$ Gyr and
$\langle \rm Z_{Samp}-Z_{Mod}\rangle = -3.7\times 10^{-9}$, with standard deviations of $1.9$ Gyr and $7\times10^{-4}$
in age and metallicities respectively. We wanted to further investigate whether the dispersion of the sampled ages and
metallicities can be used as an indicator of the real error in these quantities. The upper panel of
Figure~\ref{fig:Errors} illustrates these differences {for $10^2$ realizations of the experiment ($\sim 5\times10^9$ stars)}.
Lower panels show the absolute differences between the real ages and metallicities and the dispersion measured over the
sampled ones. Real differences in age and metallicity were above the measured dispersion in the CMD in 26 and
28\% of cases respectively. Combining both differences, only in 15\% of the cases the star had a larger real
error than the one measured through its dispersion in the CMD. This indicates that the dispersion measured over the
sampled quantities is in fact slightly overestimating the real error present in the sampled ages and metallicities and
can be safely used as a real uncertainty estimator, in particular to reject badly sampled populations.

\section{The CMD of And II}\label{Cap:CMD}
\subsection{Ages and metallicities}\label{SubCap:2DMaps:Desity_distribution}

In Figure~\ref{fig:CMD} we show the sampled CMD of And II, with the expected ages and metallicities obtained from the
SFH. Its most important features are: a clear broad RGB sequence and a blue extended horizontal branch. The RGB is
populated by intermediate-age to old stars (2-10 Gyr), while the low-mass stars populating the HB are older than 10 Gyr.
The lack of any clear bright main sequence or blue-loop stars suggests that last star formation episodes occurred in
And II more than 2 Gyr ago.

The broad RGB appears to result from the presence of two stellar populations with different metallicity distributions
(see Figure~\ref{fig:Andromeda_II_model}). In principle, the broadening of the RGB could be also explained by a large
metallicity dispersion in a single old stellar population. However, we calculated the dispersion of $\sigma(\rm [Fe/H])
\sim 0.5$ dex, and it seems unrealistic for a single old stellar population. A similar result was obtained by
\citet{DaCosta2000} who also favoured a scenario with at least two distinct stellar populations.

\subsection{Stellar population selection}\label{SubCap:2DMaps:Desity_distribution}

We cleaned our sampled photometric list from stars with errors larger than $\sigma$(Age) = 1 Gyr in age and $\sigma$(Z)
= 0.001 in metallicity using their errors measured over the sampled CMD. This selection prevents us from mixing the two
stellar populations present in And II, allowing us a more precise study of their spatial distributions. After the
cleaning, $\sim$5400 stars remained in the final photometric list, with a typical error of $\sim 0.6$ Gyr in age,
and $2\times10^{-4}$ in metallicity.

\begin{figure}
\begin{center}
\includegraphics[scale=0.7]{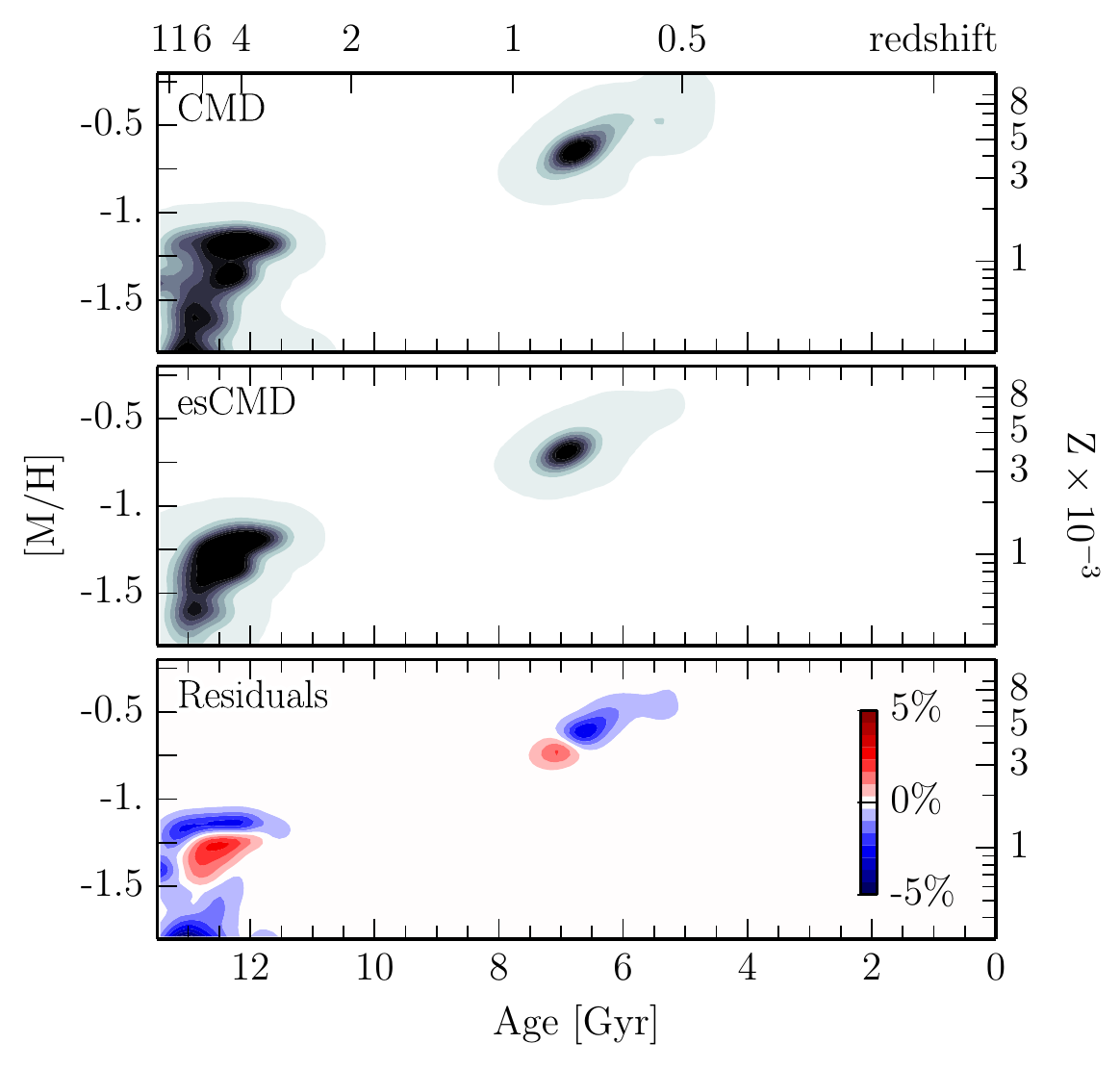}
\caption[]{Stellar populations recovered after the sampling. The top panel shows the observed CMD, the middle panel the
synthetic CMD (esCMD) and the bottom panel the percentage of the residuals in units of standard deviation.}
\label{fig:Recovered_Ages_Z}
\end{center}
\end{figure}

Figure~\ref{fig:Recovered_Ages_Z} shows the recovered age and metallicity distributions for
the observed and the synthetic CMDs. This must not be confused with the SFH of the galaxy, since it does not account for
the stars that had enough time to evolve completely and have disappeared. Both distributions are consistent,
with only marginal differences of less than 5\% in the age-metallicity plane.
This good correlation between the WFOVP CMD and the esCMD indicates two things.
First, that the assumed average SFH can be approximately considered as the SFH of the whole galaxy.
Second, that the assumed distance modulus and reddening values for the WFOVP are consistent.

{Two clearly distinct stellar populations were recovered after the cleaning: stars older than 11 Gyr and stars with ages
between 10 and 5 Gyr. These two populations are different in the Age-Metallicity plane by more than 2$\sigma$ of their distribution. Even when taking into account just the metallicity distribution of both populations, there is a difference of $\sim 1.5\sigma$ between them. These differences are compatible with the double RGB sequence present in the galaxy \citep{Weisz2014}, and indicate that both stellar populations are in fact chemically different.}

The stars in the cleaned photometry catalog were then divided into two stellar populations depending on their age
and metallicity. The first one is composed of 975 stars with ages in the range 5 Gyr $\leq$ Age $\leq$ 9 Gyr and
metallicities 0.002 $\leq$ $Z$ $\leq$ 0.01. The second group contains 4385 old stars, with ages in the range 11
Gyr $\leq$ Age, and Z $\leq$ 0.0017. This gives us the final fractions of old and intermediate-age stars present in
our photometry of $\sim80$ and $\sim20$\%, respectively.

\section{Spatial distribution of the stars}\label{Cap:2DMaps}
\subsection{Surface density maps}\label{SubCap:2DMaps:Desity_distribution}

\begin{figure*}
\begin{center}
\includegraphics[scale=0.7]{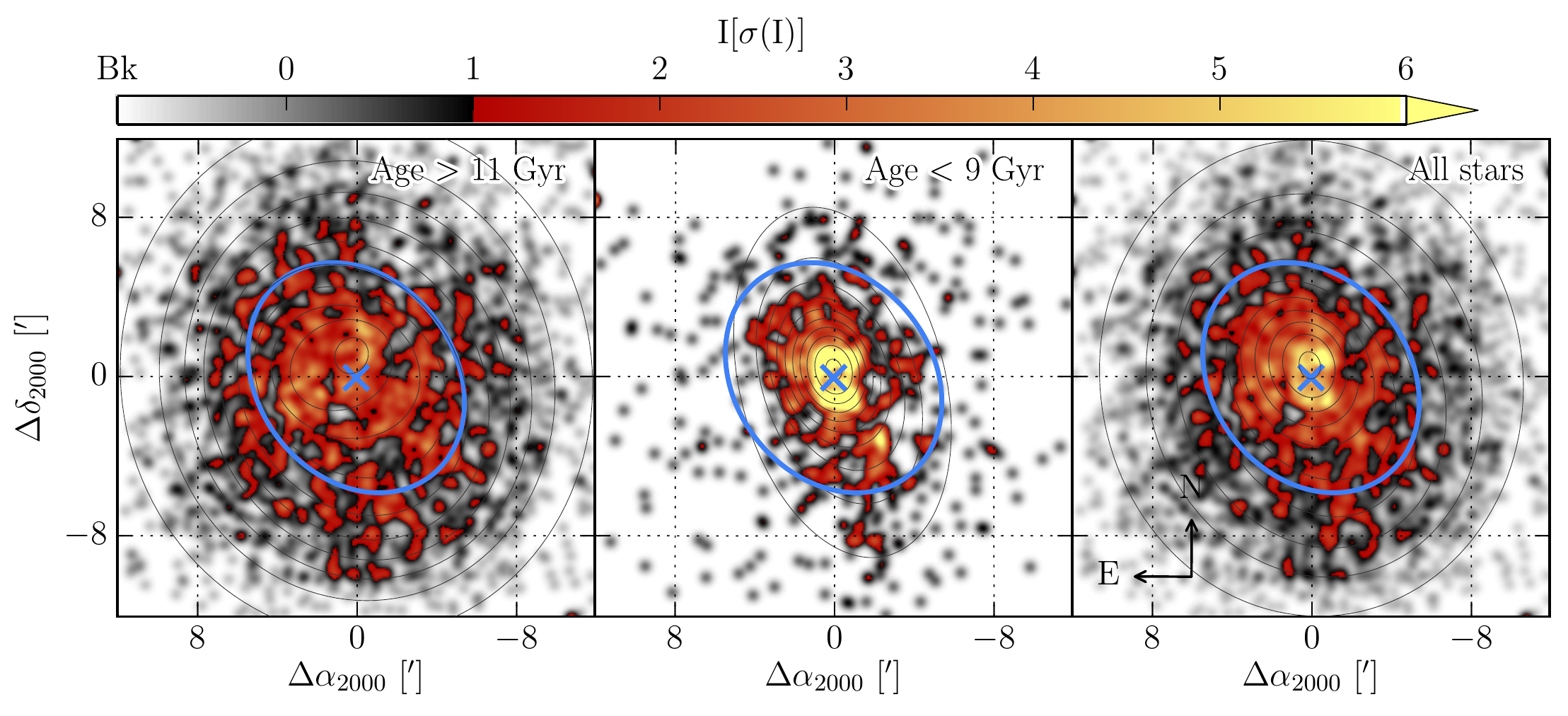}
\caption[]{Spatial distribution maps for the young and old stellar population. The colour scale is common for all
panels and indicates the concentration of stars normalized to the number of stars in each population. Levels are in
units of standard deviation for each population. Black narrow-lined ellipses represent the elliptical
boundaries obtained for deriving the radial density profiles (see Section~\ref{Cap:Radial_Profiles}). The half-light
radius of And II ($r_h = 6.2^\prime$) is represented by the blue ellipse. The centre of the galaxy is shown by a blue
cross (see Table~\ref{tab:And_II_final}).}
\label{fig:Density_Maps}
\end{center}
\end{figure*}

In order to study the spatial distribution of both stellar populations, we computed their surface density maps. These
were obtained using $512\times512$ pixel, two-dimensional histograms of the spatial distribution of the stars in the
two populations. The histograms were then convolved with a Gaussian kernel of 3-pixel width to enhance the most
important features avoiding stochastic noise. They are shown in Figure~\ref{fig:Density_Maps}.

The spatial distributions of both populations clearly differ. Old stars spread uniformly up to well beyond the
half-light radius ($r_h$) of the galaxy. Their distribution is smooth and roundish. Young stars, on the other hand, are
much more concentrated and their distribution is more elongated along the north-south direction. We also noticed that
the highest surface density point does not coincide with the barycentre of the distribution. The galaxy is also
elongated towards the south-west direction.

\begin{figure*}
\begin{center}
\includegraphics[scale=0.7]{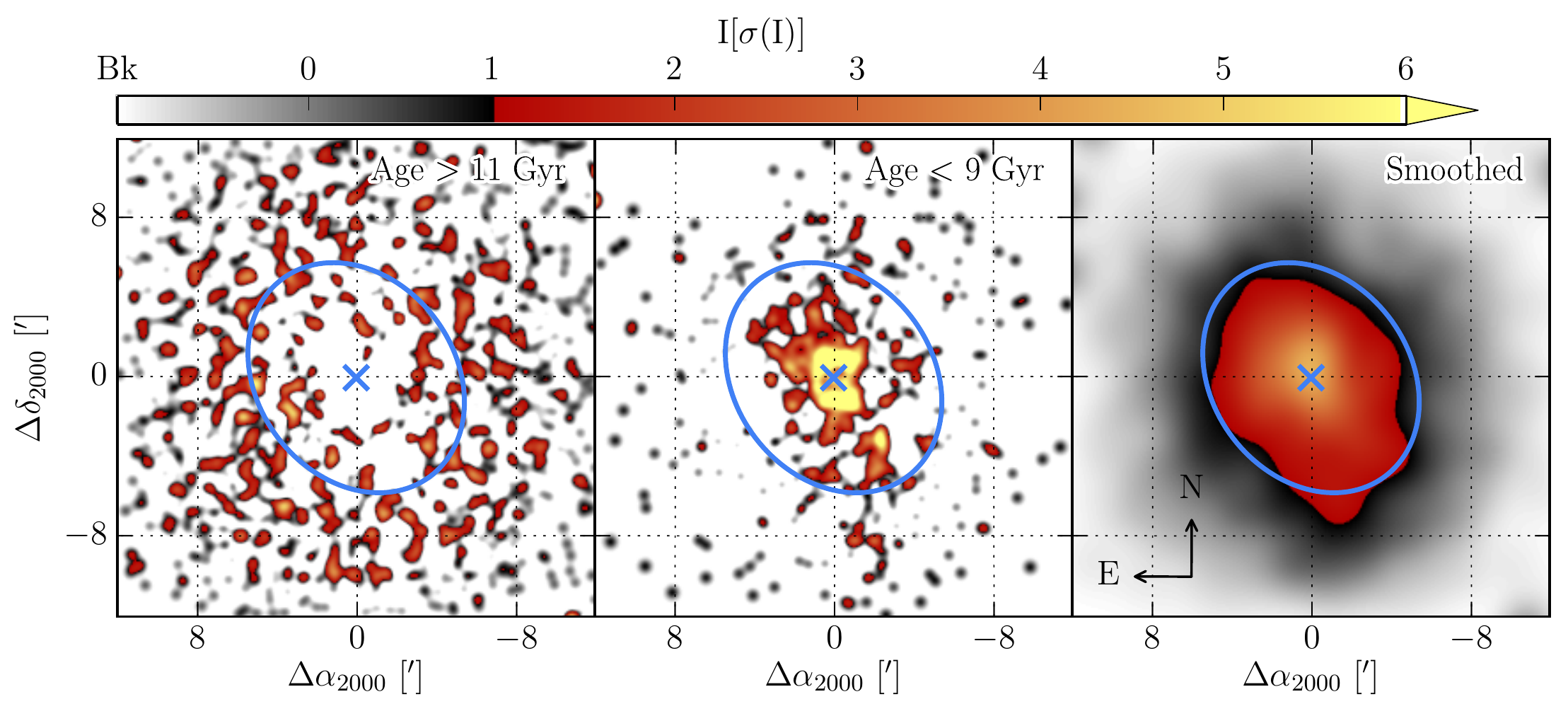}
\caption[]{Unsharp-masked version of Figure~\ref{fig:Density_Maps}.}
\label{fig:Density_Maps_Subtracted}
\end{center}
\end{figure*}

In order to enhance any possible stellar overdensity or feature present in both populations, we have computed the
unsharp-masked surface density maps from the density maps presented in Figure~\ref{fig:Density_Maps}. These were
obtained by subtracting a strongly smoothed (with a Gaussian kernel of 15 pixels) surface density map of the whole clean
sample from the young and old populations maps, all of them previously normalized to their average density. The
unsharp-masked surface density maps are shown in Figure~\ref{fig:Density_Maps_Subtracted}.

Differences between both populations are more clearly visible after subtracting the average distribution. Young stars
concentrate in the centre of the galaxy, forming a stellar density bump which extends up to $\sim2^\prime$ in
galactocentric radius ($\sim 300$ pc). This overdensity was previously noticed by \citet{McConnachie2007} as a bump in
the stellar radial density profile. In our maps, on the other hand, it appears that the distribution of the young stars
is clumpy, forming several small clusters ($r \lesssim 100$ pc) around the galaxy centre. The low density values of old
stars seen in the central regions of the galaxy are consistent with the significant contribution from young stars
to the surface density in these regions.

\begin{figure}
\begin{center}
\includegraphics[scale=0.7]{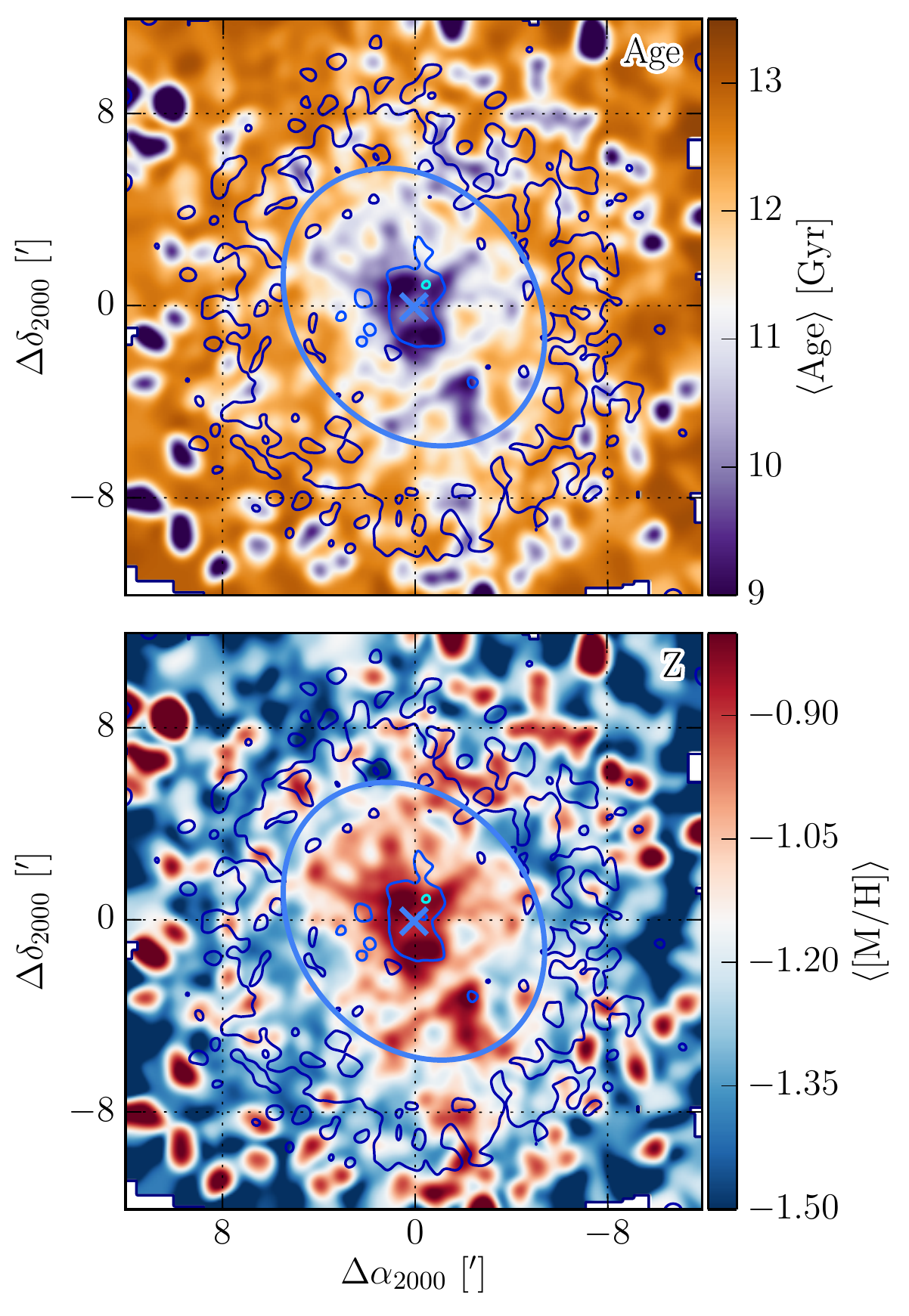}
\caption[Averaged distributions maps]{Mean distribution maps of the sampled CMD. Colour scales correspond to the average
age (top) and metallicity (bottom). Contours represent the concentration of stars in units of standard deviation with
respect to the average. Other symbols and lines are the same as in Figure~\ref{fig:Data}.}
\label{fig:Age_Map}
\end{center}
\end{figure}

Using all stars available in our cleaned photometric catalog, we have computed maps showing the average age and
metallicity distribution in the galaxy. The procedure we followed is similar to the one used for creating
Figure~\ref{fig:Density_Maps}. We computed the averaged maps as the error-weighted average of age and
metallicity of the stars, in a grid of $512\times512$ pixel in right ascension and declination. These maps are shown
in Figure~\ref{fig:Age_Map}.

The results are consistent with those of Figure~\ref{fig:Density_Maps}. In the age panel, a decrease in the average age
of the stars can be seen in the core of the galaxy. This is compatible with an increment in the average
metallicity of the galaxy following the distribution of the young stars. In general, both maps are quite irregular
which may be related to the rather coarse distribution of the young stars. We also found a drop in the average age of
the stars located approximately at one core radius at the south-west optical major axis of the system. The cause of
this is the presence of a clump of young stars, which clearly stand out over the average stellar density at these
galactocentric distances. This clump forms a part of a larger structure of young stars extending more than $10^\prime$
in galactocentric radius ($\sim 2000$ pc) towards the south, south-west direction. The origin of this structure is not
clear, but in principle we would not expect such feature to be present in a relaxed system in dynamical equilibrium.

\subsection{Ellipse fitting}\label{Cap:Ellipse}

{We fitted ellipses to the isopleths of surface density maps of the two populations
(see Figure~\ref{fig:Density_Maps}). Free parameters for the fitting were the centre, the position angle (PA),
and both semi-axis lengths of each ellipse. Apart from the fitting error itself, other sources of error should
be taken into account. Particularly we found that the obtained results can substantially change when using
different sampling resolutions for creating such surface density maps or due to the size of the kernel adopted to
smooth them. In order to account for these variations, we measured the stellar surface density over 100 different
grids with resolutions spanning from 150$\times$150 pixels to 4092$\times$4092 convolved with 50 different
Gaussian kernels of sizes varying between 1/25 and 1/10 of the map size. In addition, we used different
isodensity steps in each map in order to obtain its isopleths. Steps of 10, 7.5, 5, and 2.5\% of the total
variation of surface density across each map were used. In total, we applied 20000 different combinations per stellar
population. In order to derive the final ellipse parameters at a given radius, we applied a moving average
over all the realizations of the experiment for each population, obtaining the expected average value and its
dispersion as a function of the semi-major axis length ($a$).}

We found that the WFOVP field is affected by numerous foreground stars. The large apparent brightness of these saturate
the CCD within a certain radius making it impossible to recover And II stars. The first consequence of these gaps in the
photometry is the apparent lack of stars in the saturated areas, resulting in unrealistic stellar underdensities that
could affect both the isopleths calculation and the measured stellar densities within the elliptical regions. We
corrected for this by masking the saturated areas in the smoothed density maps and by subtracting the equivalent
areas from the corresponding annular regions. We used the point source catalog from the Sloan Digital Sky Survey
\citep{SLOANDR7} to remove and mask up to 2718 foreground stars up to the magnitudes $g = 19$. The final and corrected
ellipse sets are shown in Figure~\ref{fig:Density_Maps}. The parameters of ellipses fitted at $r_h$ are listed in
Table~\ref{tab:Ellipses_promedio}. The columns of this Table provide the following information:

\begin{itemize}
\item Column 1: population;
\item Column 2, 3: coordinates of the centre of the ellipse fitted at $r_h$ ($\alpha_{2000}$, $\delta_{2000}$);
\item Column 4: ellipticity $\epsilon = 1 - b/a$, where $b$ is the
  semi-minor axis and $a$ is the semi-major axis;
\item Column 5: position angle, PA, measured as the angle between the semi-major axis and the direction to the north.
\end{itemize}

\begin{table*}
  \caption{Parameters of the ellipses at $r_h$ for each stellar population.}
  \label{tab:Ellipses_promedio}
  \begin{tabular}{@{}lcccc}
    \hline
    \hline
    Populatio & $\alpha_{2000}$  & $\delta_{2000}$ & $\epsilon$ & PA\\
        &                  &                 &          & [deg]\\
    \hline
    Age $>$ 11 Gyr  &  $\rm 1^h$ $\rm 16^m$ $\rm 27.90^s \pm 0.08^s$  & $33^{\circ}$ $25^\prime$ $21^{\prime\prime} \pm 2^{\prime\prime}$  &  $0.17\pm0.02$ & $38\pm3$ \\
    Age $<$  9 Gyr  &  $\rm 1^h$ $\rm 16^m$ $\rm 26.12^s \pm 0.08^s$  & $33^{\circ}$ $25^\prime$ $47^{\prime\prime} \pm 3^{\prime\prime}$  &  $0.33\pm0.05$ & $24\pm2$ \\
    Total           &  $\rm 1^h$ $\rm 16^m$ $\rm 27.19^s \pm 0.09^s$  & $33^{\circ}$ $25^\prime$ $32^{\prime\prime} \pm 3^{\prime\prime}$  &  $0.22\pm0.03$ & $31\pm3$ \\
    \hline
  \end{tabular}
\end{table*}

\begin{figure}
\begin{center}
\begin{flushleft}
\includegraphics[scale=0.7]{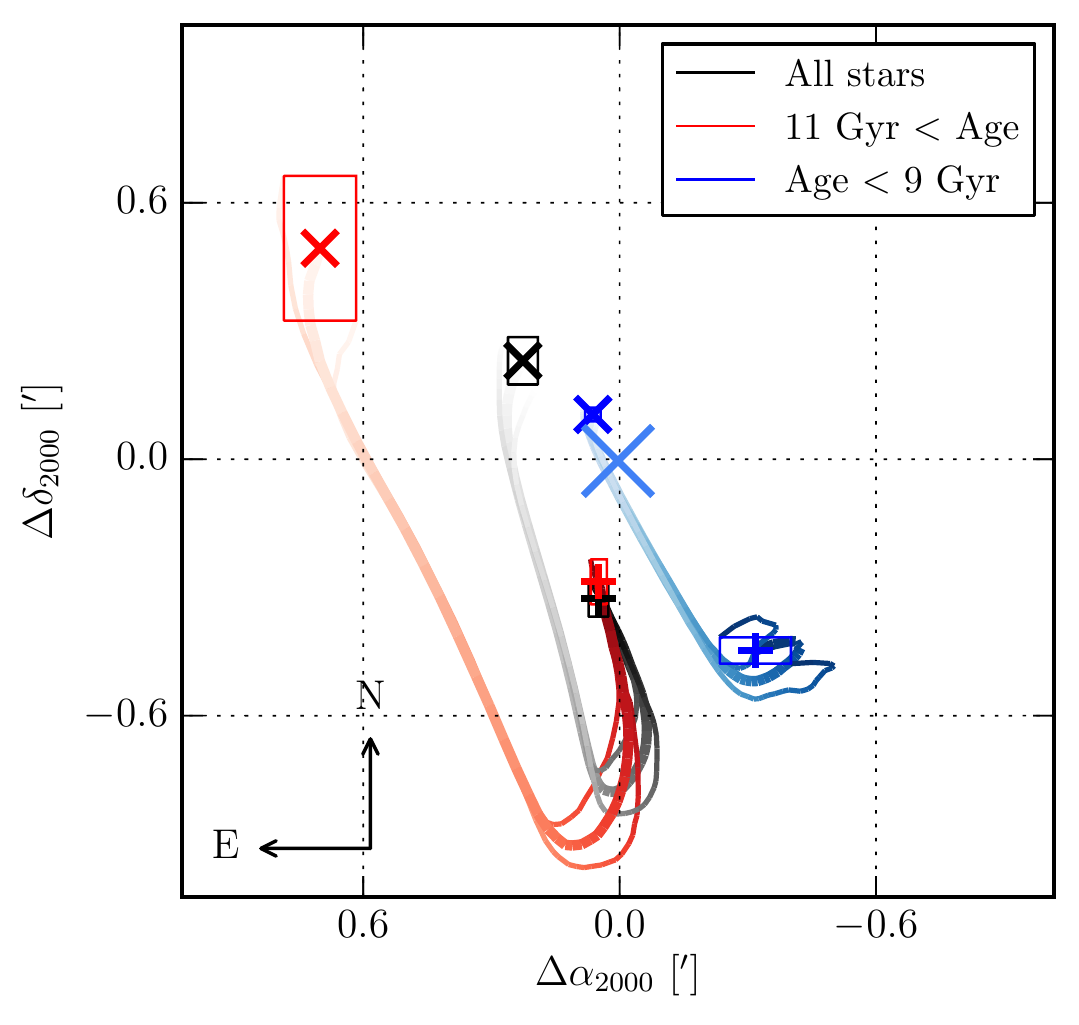}
\includegraphics[scale=0.7]{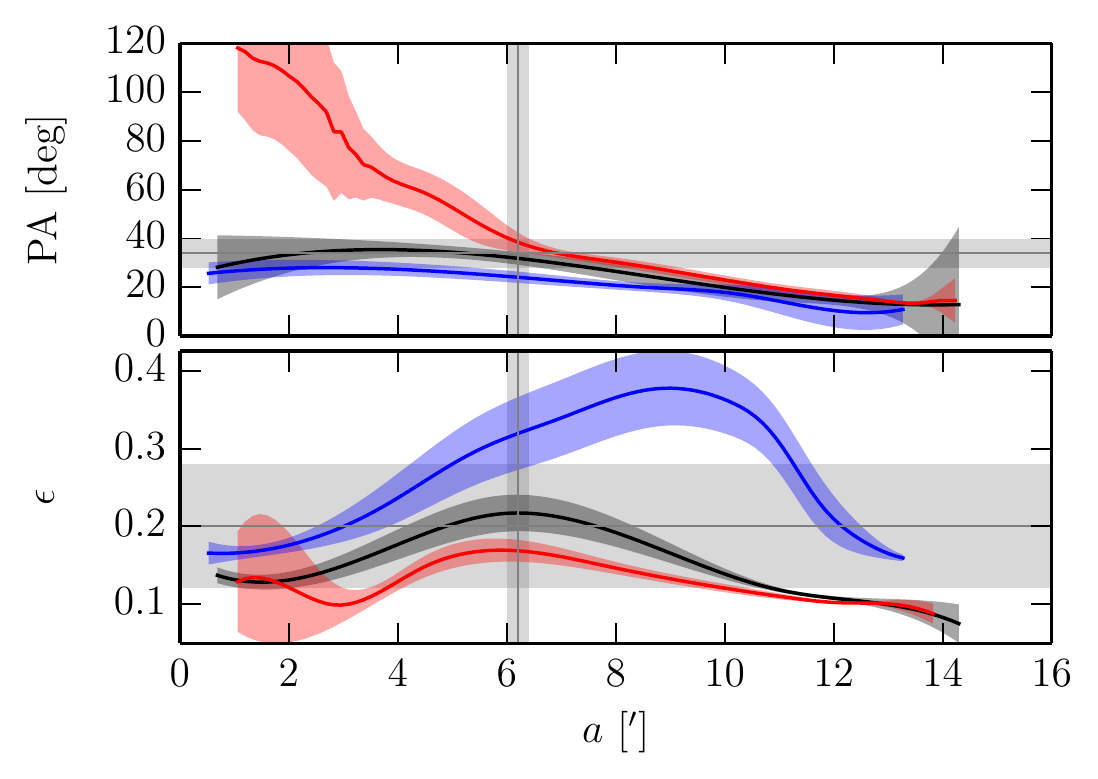}
\end{flushleft}
\caption[Fitted ellipse parameters as a function of radius]{Evolution in the shape of the ellipses as a function of
their semi-major axis length ($a$). Top panel: the position of the centre of the ellipses ($\Delta\alpha_{2000}$,
$\Delta\delta_{2000}$) with respect to the adopted centre of And II (Table~\ref{tab:And_II_final}). {The position of the
smallest ellipses are marked with cross symbols in red and blue for the old and the intermediate-age populations,
respectively. The black cross marks the centre of the smallest ellipse fitted to our whole cleaned sample.
Plus symbols of the same colour show the
position of the centres of the largest ellipse fitted to each population. Thick-lined curves indicate the evolution in
the position of the centres with $a$, while their $1\sigma$ errors are shown by thin curves. The colour shades of the
curves indicate the growth with $a$. Hollow boxes show the error for the largest and smallest ellipses of each population.}
Bottom panels:
the position angle (PA) and ellipticity ($\epsilon$) as a function of $a$. Shaded areas indicate $1\sigma$ error for
each curve. The half-light radius of the galaxy ($r_h = 6.2\pm0.2 ^\prime$) is marked with a vertical grey-shaded area, 
while the PA and $\epsilon$ measured at this radius are shown by the horizontal grey-shaded areas in the respective panels
\citep{McConnachie2012}.}
\label{fig:Ellipses_evolution}
\end{center}
\end{figure}

The properties of the ellipses vary differently with $a$ for each population, roughly tracing the spatial
distribution of each population at different galactocentric distances. Figure~\ref{fig:Ellipses_evolution} shows the
variations in the ellipse centres, PA and $\epsilon$ as a function of the galactocentric distance, $a$.

The two populations do not share the same baricentre. At small galactocentric distances, the centres of both
distributions are located in the north-east direction with respect to the centre of the galaxy. The old population
shows the largest displacement in the position of its centre, moving by more than 1 arcmin from its original position to
its final one, located at $\sim3$ arcmin to the south from the galaxy centre. This is related to the elongation
of the galaxy at large radii towards the south-west direction (see Figure~\ref{fig:Density_Maps} and
\ref{fig:Density_Maps_Subtracted}). Younger stars also suffer from this effect at large radii, but less significantly.
The separation of each population with respect to the centre of the total sample indicates how numerous are the
stars of each population with respect to the total depending on $a$. The intermediate-age population clearly dominates
in the centre of the galaxy, but its importance decreases rapidly with the distance to the centre, and the old
population is the dominant one at radii larger than $r_h$. This can be seen also in the lower panels of
Figure~\ref{fig:Ellipses_evolution}: the shape of the total stellar population {converges with the one of the old
population when going outwards.}

{Ellipticities of both populations also evolve in a completely different way. The old population seems quite round at all
galactocentric distances, with $\epsilon$ ranging between $\sim 0.1$ for $a \gtrsim 11 ^\prime$ up to $\sim 0.17$ at
$a \gtrsim 11 ^\prime$. A different trend is seen for the younger stars. These gather in the centre of And II forming a quite
round distribution which changes into a very elongated one ($\epsilon \sim 0.4$) for radii larger than $r_h$. The
PAs of the two populations also differ. At small radii both populations show almost perpendicular PAs. The old
population PA goes from $\sim 120$ to $\sim 20$ deg.} At small radii, this is almost perpendicular to the average
value for the galaxy PA, 34 deg. Younger stars, on the other hand, extend along the north-south direction with smaller
PA values, between 30 and 10 deg. These differences in the shape of both populations may indicate the presence of
dynamically decoupled components in And II. In Table~\ref{tab:And_II_final} we list the structural parameters derived
for the whole sample of stars. These were calculated as the error weighted average of these parameters for the whole
ellipse set derived for the whole sample of stars without any age or metallicity error cut (24562 stars).

\begin{table}
\centering
  \caption{Main structural parameters at $r_h$ for the whole sample of stars, without any cut in the age or metallicity
  errors (24562 stars).}
  \label{tab:And_II_final}
  \begin{tabular}{@{}lc}
    \hline
    \hline
    Quantity & Value \\
    \hline
    RA, $\alpha$ (J2000.0) &  $\rm 1^h$ $\rm 16^m$ $\rm 26.94^s \pm 0.06^s$ \\
    Dec., $\delta$ (J2000.0) & $33^{\circ}$ $25^\prime$ $39.7^{\prime\prime} \pm 0.9^{\prime\prime}$ \\
    Ellipticity, $\epsilon$ & $0.23\pm 0.03$ \\
    PA $[^\circ]$ & $36\pm2$ \\
    \hline
  \end{tabular}
\end{table}

\subsection{The radial density profiles}\label{Cap:Radial_Profiles}

{We obtained the radial density profiles of the two stellar populations over the entire set of ellipses
defined in the previous section. These were calculated as the error-weighted average of all the combinations
defined in the ellipses sets. A box-car moving average was applied over the 20000 different profiles derived
for each population, obtaining their continuous average stellar density and its error as a function of radius.
We adopted the semi-major axis length ($a$) as galactocentric radius. Errors account for uncertainties
in the ellipse fitting procedure, including Poissonian error from counting the sources and from the contaminant
subtraction.}

We fitted three models to the obtained radial density profiles of the stars: a Plummer sphere model
\citep{Plummer1911}, a King model \citep{King1962, King1966} and a S\'ersic profile \citep{Sersic1968}.

The Plummer and King are dynamical models for self-gravitating stellar systems. The Plummer sphere model is
parametrized by the total stellar mass of the system, $M$, and a characteristic scale length $\alpha$. Assuming the
same mass for all stars, $M$ can be replaced by the total number of stars $N_\star$ in the system. Therefore, the
number of stars per unit of area as a function of the radius, $I(r)$, can be expressed as: $$I(r) =
\frac{\alpha^2N_\star}{\pi(r^2 + \alpha^2)^2}$$.

The slightly more complicated King model, uses two different scale lengths to better reproduce truncated profiles. It
can be expressed as: $$I(r) = I_0\left[\frac{1}{[1+(r/r_{\rm c})^2]^{1/2}} - \frac{1}{[1+(r_{\rm t}/r_{\rm
c})^2]^{1/2}}\right]^2$$ where $I_0$ is the central stellar surface density, $r_{\rm c}$ the core radius, and $r_{\rm
t}$ is the tidal radius.

The last fitted model, the S\'ersic profile, is a purely empirical model of the form $$I(r) = I_0
\exp\left[-\left(\frac{r}{r_{\rm c}}\right)^{1/n}\right]$$ in which, as in the King model, $I_0$ and $r_{\rm
c}$ stand for the central density and the core radius, while $n$ is a shape parameter, the so-called S\'ersic index.

\begin{figure*}
\begin{center}
\includegraphics[scale=0.8]{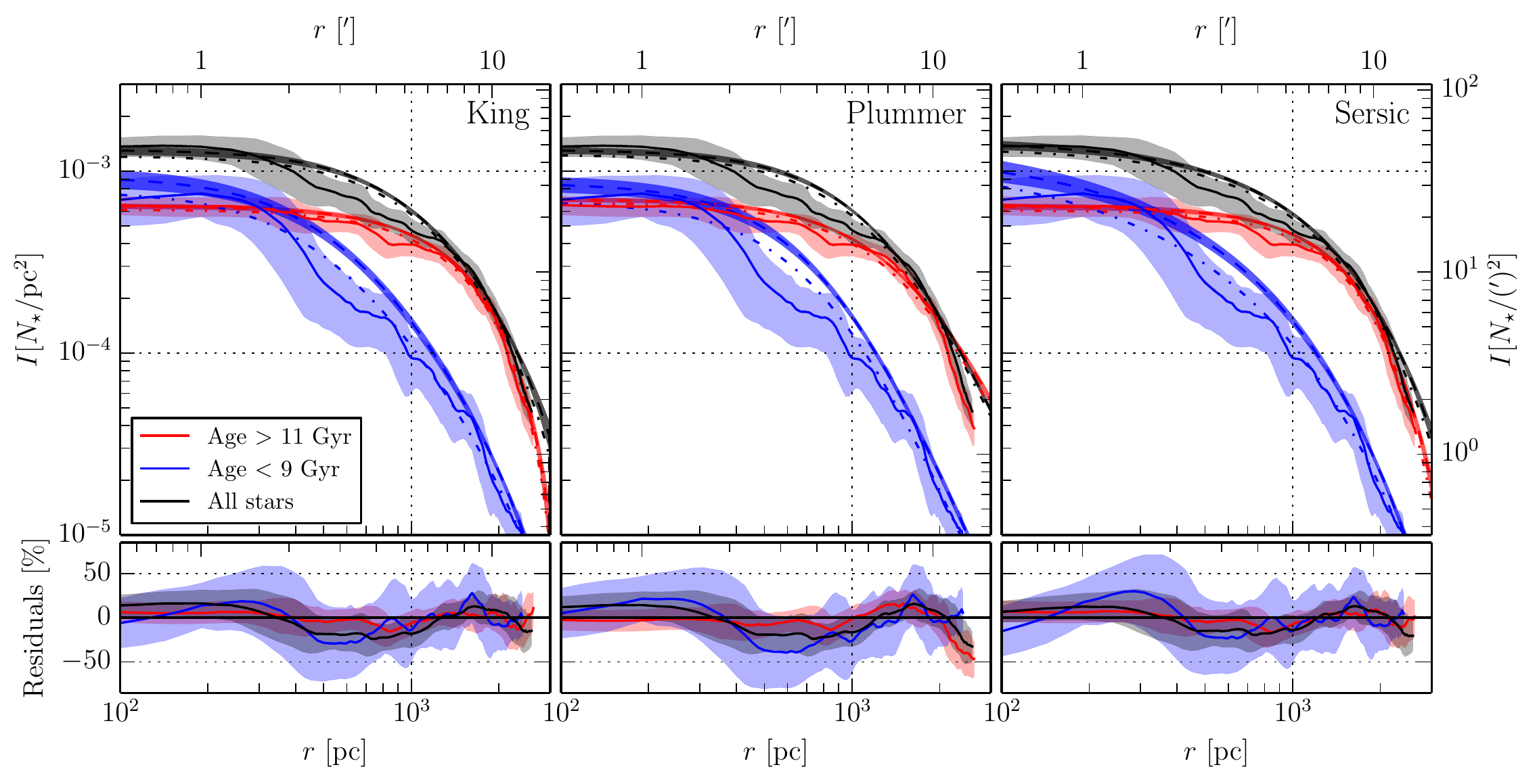}
\caption[Radial surface density profiles]{Radial surface density profiles for different populations. {Continuous lines
show the stellar surface density profiles while their errors are indicated by light shaded areas.}
The three fitted models: King, S\'ersic and Plummer are
shown in the three panels from the left to the right and labelled accordingly. The best fits of these models found using
the least squares approach are shown with {dotted-dashed} lines, while those obtained through Monte Carlo inference are
shown with dashed lines. Darker shaded areas indicate the $\pm1\sigma$ of the goodness of fit with Monte Carlo. Bottom
panels show the residuals from the fit using least squares.}
\label{fig:Radial_Profiles}
\end{center}
\end{figure*}

In order to account for possible errors, we derived the parameters of each model through extensive Monte Carlo
experiments. In each experiment, a random shift is introduced in each point of the observed radial profile. These
shifts are randomly generated for each point from a normal probability distribution function centred on the observed
values of the galactocentric distance and star counts using their associated errors as its standard deviation. The three
models are then fitted to this profile using error-weighted least squares. The error of each model parameters {are}
determined from the covariance matrix of the fit and the residual variance. The experiment is performed $1\times10^{6}$
times, storing all possible results and their associated errors. The median of all the possible values and its standard
deviation was adopted as the final value for each fitting parameter. The radial profiles for both stellar populations
and their fitted models are shown in Figure~\ref{fig:Radial_Profiles}. The fitted parameters for the King, Plummer and
S\'ersic models are listed in Tables \ref{tab:King_model}, \ref{tab:Plummer_model} and \ref{tab:Sersic_model},
respectively. The format of the three tables is similar, except for the listed parameters:

\begin{itemize}
\item Column one: population;
\item Column second to penultimate: best-fitting parameters;
\item Last column: $\chi^2$ resulting from the fitting.
\end{itemize}

\begin{table*}
  \caption{Fitted parameters for the King model for each population.}
  \label{tab:King_model}
  \begin{tabular}{@{}lccccccc}
    \hline
    \hline
    Population & \multicolumn{7}{c}{King parameters} \\
            & \multicolumn{2}{c}{$I_0$} & \multicolumn{2}{c}{$r_{\rm c}$}  & \multicolumn{2}{c}{$r_{\rm t}$}                                 & $\chi^2$ \\
            & $\rm[pc^{-2}]\times10^{-3}$       & $\rm[(^\prime)^{-2}]$ & [pc] &  [$^\prime$] & [pc] &  [$^\prime$]  &        \\
    \hline
Age $>$ 11 Gyr & $0.7\pm0.3$ & $25\pm9$ & $3100\pm400$ & $16\pm2$ & $3110\pm90$ & $16.4\pm0.5$ & 1.36  \\
Age $<$ 9 Gyr & $1.1\pm0.1$ & $41\pm4$ & $430\pm50$ & $2.3\pm0.3$ & $6000\pm2000$ & $32\pm8$ & 2.58  \\
All stars & $2.5\pm0.3$ & $90\pm10$ & $1300\pm100$ & $7.1\pm0.7$ & $4000\pm300$ & $21\pm2$ & 3.40  \\
    \hline 
  \end{tabular}
\end{table*}

\begin{table*}
  \caption{Fitted parameters for the Plummer model for each population.}
  \label{tab:Plummer_model}
  \begin{tabular}{@{}lcccccc}
    \hline
    \hline
    Population & \multicolumn{4}{c}{Plummer parameters}\\
    & $N_\star$ & \multicolumn{2}{c}{$\alpha$} & $\chi^2$ \\
    &     $\times10^{3}$        & [pc] &  [$^\prime$]\\
    \hline
Age $>$ 11 Gyr  & $7.1\pm0.2$ & $1840\pm40$ & $9.7\pm0.2$ & 3.56  \\
Age $<$ 9 Gyr & $1.63\pm0.03$ & $880\pm10$ & $4.61\pm0.06$ & 1.41  \\
All stars & $8.2\pm0.1$ & $1460\pm20$ & $7.7\pm0.1$ & 1.62  \\
    \hline
  \end{tabular}
\end{table*}

\begin{table*}
  \caption{Fitted parameters for the S\'ersic model for each population.}
  \label{tab:Sersic_model}
  \begin{tabular}{@{}lcccccccccc}
    \hline
    \hline
    Population     & \multicolumn{6}{c}{S\'ersic parameters}\\
                   & \multicolumn{2}{c}{$I_0$}   & \multicolumn{2}{c}{$r_{\rm c}$} & $n$ & $\chi^2$\\
                   & $\rm[pc^{-2}]\times10^{-3}$ & $\rm[(^\prime)^{-2}]$ & [pc] &  [$^\prime$] &  &  \\
    \hline
Age $>$ 11 Gyr  & $0.60\pm0.04$ & $21.8\pm0.6$ & $1630\pm40$ & $8.59\pm0.07$ & $0.42\pm0.04$ & 0.97  \\
Age $<$ 9 Gyr & $1.2\pm0.7$ & $42\pm3$ & $350\pm30$ & $1.8\pm0.1$ & $1.35\pm0.25$ & 1.53  \\
All stars & $1.3\pm0.1$ & $48\pm2$ & $1100\pm50$ & $5.8\pm0.2$ & $0.74\pm0.07$ & 1.90  \\
    \hline
  \end{tabular}
\end{table*}

The results are qualitatively consistent between the models, although the King model always requires larger scale
radii for all populations. The S\'ersic profile provides the best fitting model overall, followed by the King and the
Plummer sphere profiles. In the following, we will refer to S\'ersic scale lengths to give a consistent
discussion of the results.

The profile of the old population is quite unusual. It extends up to relatively large distances ($r_{\rm c} =
1630\pm40$ pc) with an almost constant density. Beyond that distance, star counts experience a sharp drop. This is
reflected in the very low S\'ersic index of this population, $n = 0.42\pm0.04$, indicating that the profile is
basically truncated at large radii. This sharp truncation can be also observed in the King profile scale radii, $r_{\rm
c}$ and $r_{\rm t}$, which are almost the same for the old stars, and also in the overestimate of the stellar density
at large radii by the Plummer sphere model. Intermediate-age stars are in general worse fitted by 1-dimensional radial
models, probably due to their more irregular distribution. They are much more concentrated at small radii ($r_{\rm c} =
350\pm30$ pc), and show a much larger S\'ersic index ($n = 1.35\pm0.25$), indicating a quite extended distribution
towards large radii. {In fact, the large S\'ersic index is a consequence of the large elongation in the North to
South-West direction shown by the intermediate-age population. This elongation can be noticed in
Figures ~\ref{fig:Density_Maps} and ~\ref{fig:Density_Maps_Subtracted}.}

Our results are consistent with previous works. Using the same data set, \citet{McConnachie2007} defined four regions
in the CMD and obtained the S\'ersic parameters for their radial density profiles. Two of these regions, the red RGB
(iii), presumably composed of younger stars, and the HB (iv) composed of old stars should be comparable to our
intermediate-age subsample and to our old subsample, respectively. For iii they obtained $r_{\rm c} = 2.18\pm0.78$
arcmin and $n = 1.03\pm0.29$, while for iv the values were $r_{\rm c} = 10.00\pm8.48$ arcmin and $n = 0.30\pm0.11$.
This is compatible {within $1\sigma$} with our results for the intermediate-age and old populations, respectively.

\begin{figure}
\begin{center}
\includegraphics[scale=0.7]{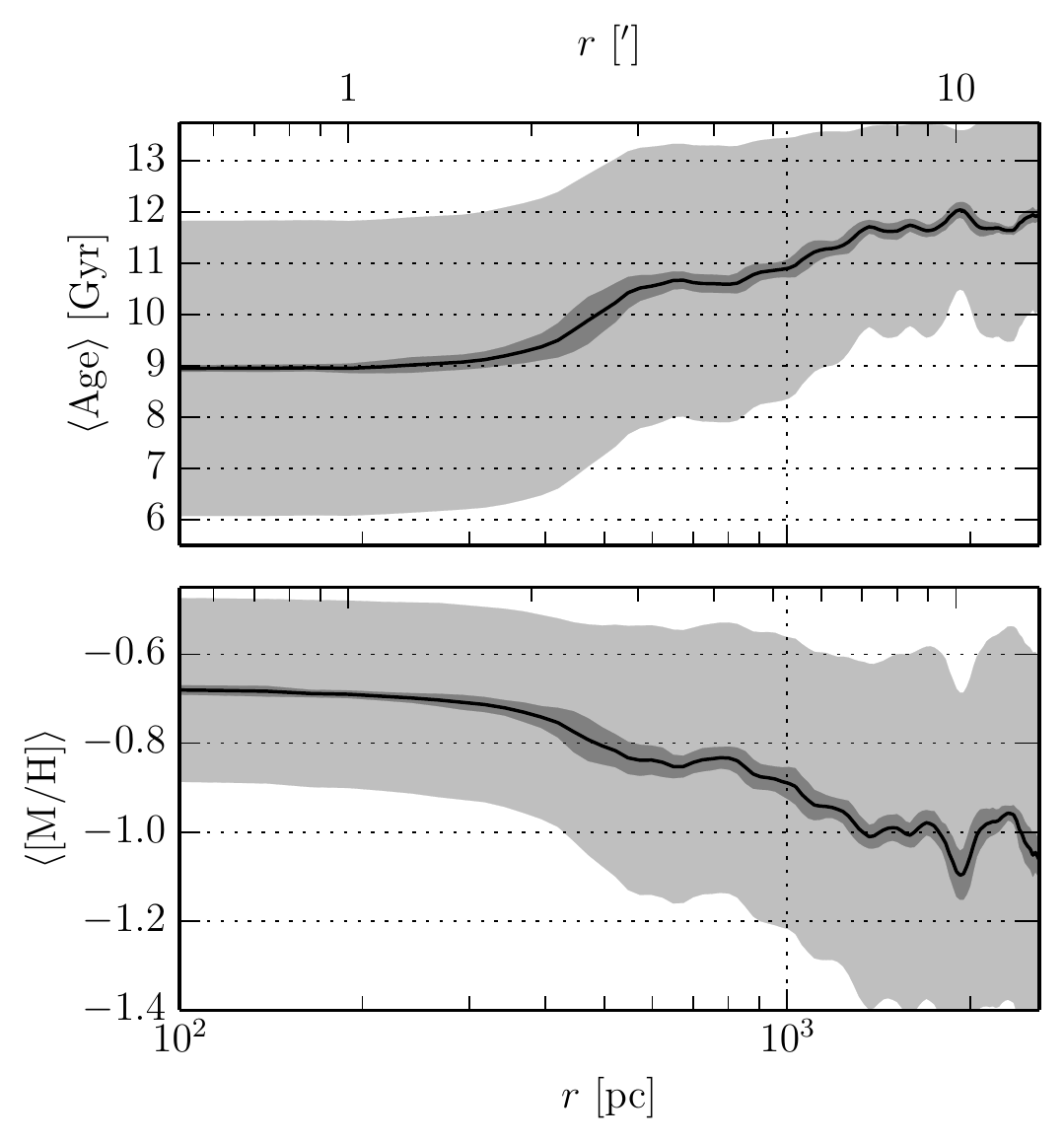}
\caption[Radial surface density profiles]{{Average age (top panel) and metallicity (bottom panel) as a function of radius.
Dark shaded areas indicate the error of the average, while lighter areas indicate the actual $\pm1\sigma$ of the age and
metallicity distributions at a given radius.}}
\label{fig:Radial_Age}
\end{center}
\end{figure}

Following a similar procedure as the one used for deriving the stellar radial density profiles, we have measured
the averaged age and metallicity profiles from the cleaned total sample. These were calculated using a moving average of
the age and metallicities of the stars within the elliptical regions defined in the previous section. The profiles are shown
in Figure~\ref{fig:Radial_Age}. Both curves seem rather constant up to $r \sim 400$ pc. After that, an increment in the
average age, as well as a drop in the average metallicity of the galaxy can be observed. This is related to the quite
flat distribution of intermediate-age stars in the central regions of the galaxy.

\section{Chemo-kinematics of And II}\label{Cap:Spectra}
\subsection{And II rotation signal}\label{Cap:Rotation_Signal}

In the previous sections, we have analyzed the distribution of the stellar content in And II using the photometry of
the stars. The old and intermediate-age stellar populations are dramatically different in their shapes and
concentration. Assuming that both populations are in dynamical equilibrium in the global potential of the galaxy it
is very likely that their dynamical properties also differ. Another possibility is that one or even both
populations are not in equilibrium, and therefore they do not have to share dynamical properties at all, 
however in this case they would not be expected to survive for much longer than the dynamical time. In
addition, And II shows a strong prolate rotation around its optical major axis. It is worthwhile to investigate whether
the odd kinematics of And II and the distribution of its stars are related.

\begin{figure*}
\begin{center}
\includegraphics[ scale=0.7]{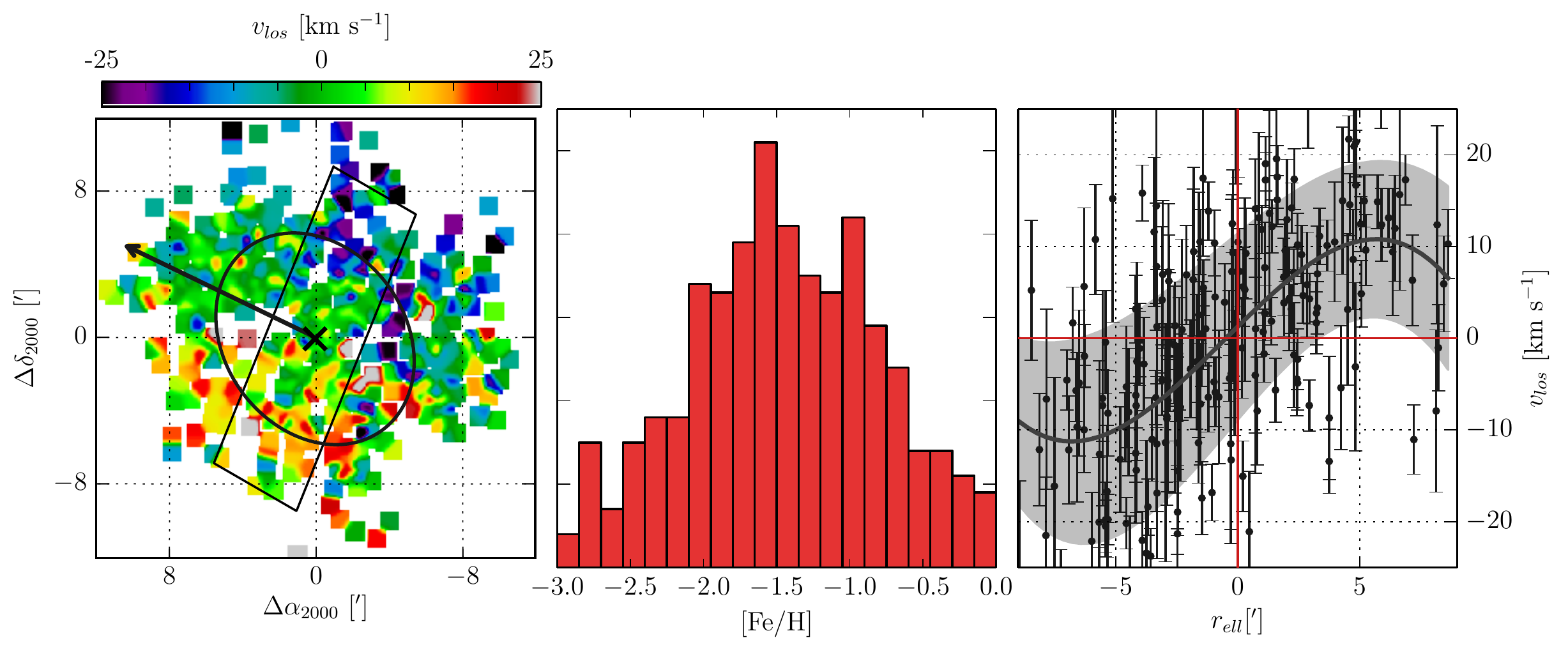}
\caption[Stars used for the chemo-kinematic analysis]{The map of line-of-sight velocities, \vlos, for the 544 stars
from the \citet{Ho2012} sample. The left panel shows the \vlos\ as a function of position. The half-light radius,
$r_h$ is represented by the black ellipse. The middle panel shows the metallicity distribution of the stars present in
the sample of the left panel. The right panel plots the velocities of the subsample of stars lying within the stripe
marked in the left panel. The average \vlos~ measured along the stripe is shown with a grey line, while its dispersion
is represented by the shaded area. The galactocentric radii of the stars are given in elliptical radius, $r_{ellip}$,
using the parameters listed in Table~\ref{tab:And_II_final}. The arrow in the left panel shows the {projected} angular momentum,
$\mathbf{L}$, of the galaxy.}
\label{fig:rawmaps}
\end{center}
\end{figure*}

We have used the sample of RGB stars from \citet{Ho2012} to reanalyze the dynamical properties of the galaxy and to
determine whether the two stellar populations are in fact dynamically decoupled. In Figure~\ref{fig:rawmaps} we show the
line-of-sight velocity map, \vlos, for the 544 RGB stars catalogued as members of And II by \citet{Ho2012}. The black
arrow in the left hand panel indicates the projected angular momentum,
$\mathbf{L}  = \sum_{i=1}^{i=j}{\mathbf{r}_i \times m_i\mathbf{v_{ los, i}}}$, where $\mathbf{r}_i$,
$m_i$ are the position and the mass of the $i^{\rm th}$ star and $j$ is the number of stars.
For its calculation we have assumed that all stars have the same mass. The prolate rotation signal can be
seen as the angular momentum vector points roughly along the optical major axis of the galaxy. {The right panel of
Figure~\ref{fig:rawmaps} shows the velocity curve measured along the stripe shown in the left panel (perpendicular to $\mathbf{L}$). This was obtained by fitting a cubic spline to the error weighted moving average of
the \vlos~ of all the stars lying inside the stripe.} We have used this sample of stars as an input for the \beacon~software in order to detect possible
different chemo-kinematic patterns among these stars.

\subsection{What is Beacon?}\label{Cap:WhatisBEACON}

\beacon~is a code designed specifically for finding chemo-kinematic patterns in resolved stellar systems
\citep{delPino2017}. The term stellar chemo-kinematics refers to the combined information about the chemical
composition ($\rm [Fe/H]$) and the kinematics of the stars. The core of \beacon~is based on the {\sc Optics} algorithm
by \citet{Optics1999}, which is used to find groups of stars with similar positions, velocities and metallicities. Its
scientific objective is to detect possible rotation patterns or streams in resolved stellar populations. It was
successfully tested with $N$-body simulations and its capabilities and features were presented in more detail in
\citet{delPino2017}. The procedure is roughly summarized here:

We first create a state vector $\mathbf{\Phi}$, representative of our sample of stars. This vector will include the
coordinates we consider important for detecting kinematic patterns, usually RA, Dec and \vlos~of the stars, all of
them expressed in the galaxy rest frame. Therefore, the coordinates of the galaxy have to be also provided to
\beacon~in order to assign a position and \vlos~to the centre of mass (CM) of the system (RA$^{\rm CM}$, Dec$^{\rm
CM}$, ${v}^{\rm CM}_{los}$). Lastly, the metallicities of the stars will be used as the fourth coordinate, assuming that
stars showing similar kinematic patterns should also have similar chemical composition. For the systemic radial
velocity of And II, we adopted $v^{\rm CM}_{los} = 192.4 \pm 0.5$ km s$^{-1}$ \citep{Ho2012}. For the shape and
position parameters, we decided to perform an experiment using as a first guess the three different coordinates
corresponding to each stellar population found in the galaxy. The data used were taken from
Table~\ref{tab:Ellipses_promedio} for the old and intermediate-age populations, and Table~\ref{tab:And_II_final} for
the whole sample. After the process, the state vector would have the explicit form $\mathbf{\Phi} = (\theta_i, r_{ell,
i}, v_{los, i}, {\rm [Fe/H]}_i)$, where each $\theta_i$ is the angular coordinate of the $i^{\rm th}$ star with respect
to the CM and the optical minor axis of the galaxy and each ${v}_{{los},i}$ is measured with respect to the ${v}^{\rm
CM}_{los}$.

In the second step, \beacon~will find groups of stars with similar metallicities that are presumably moving around the
CM. These are groups of stars located on one side of the CM, moving towards us with \vlos$ < v^{\rm CM}_{los}$, and
their counterparts, symmetrically located on the other side of the CM and moving away with \vlos$ > v^{\rm CM}_{los}$.
Such a task requires some extra parameters defining the clustering criteria. These \textit{clustering parameters} are
the \textit{standardization method}, the \textit{uniqueness criteria} and the \textit{minimum cluster size}
\citep{delPino2017}. The \textit{standardization method} affects the relative importance of each coordinate during the
clustering. We applied a dynamical range standardization, which divides each coordinate by its range i.e. difference
between the maximum and the minimum value. The \textit{uniqueness criteria} were set to 'any element'. This will force
\beacon~to merge groups with at least one common star. Lastly, the \textit{minimum cluster size} was
set to 20. This is the minimum number of stars that a group should have. Groups with less stars than the
\textit{minimum cluster size} will be rejected or merged with other ones until all the resulting groups have more stars
than the \textit{minimum cluster size}. After the clustering process the groups are catalogued as a possible circular
or both side streams (BSS) depending on the proportion of stars which are rotating on both sides of the CM. A ratio of
3 to 1 {between the number of stars found at both sides of the CM} is the maximum allowed. More unbalanced groups are rejected. The final product of the procedure is a catalogue
of all stars found to belong to any specific chemo-kinematic patterns or BSSs.

\subsection{Multiple dynamic components}\label{Cap:WhatisBEACON}

\begin{figure*}
\begin{center}
\includegraphics[ scale=0.7]{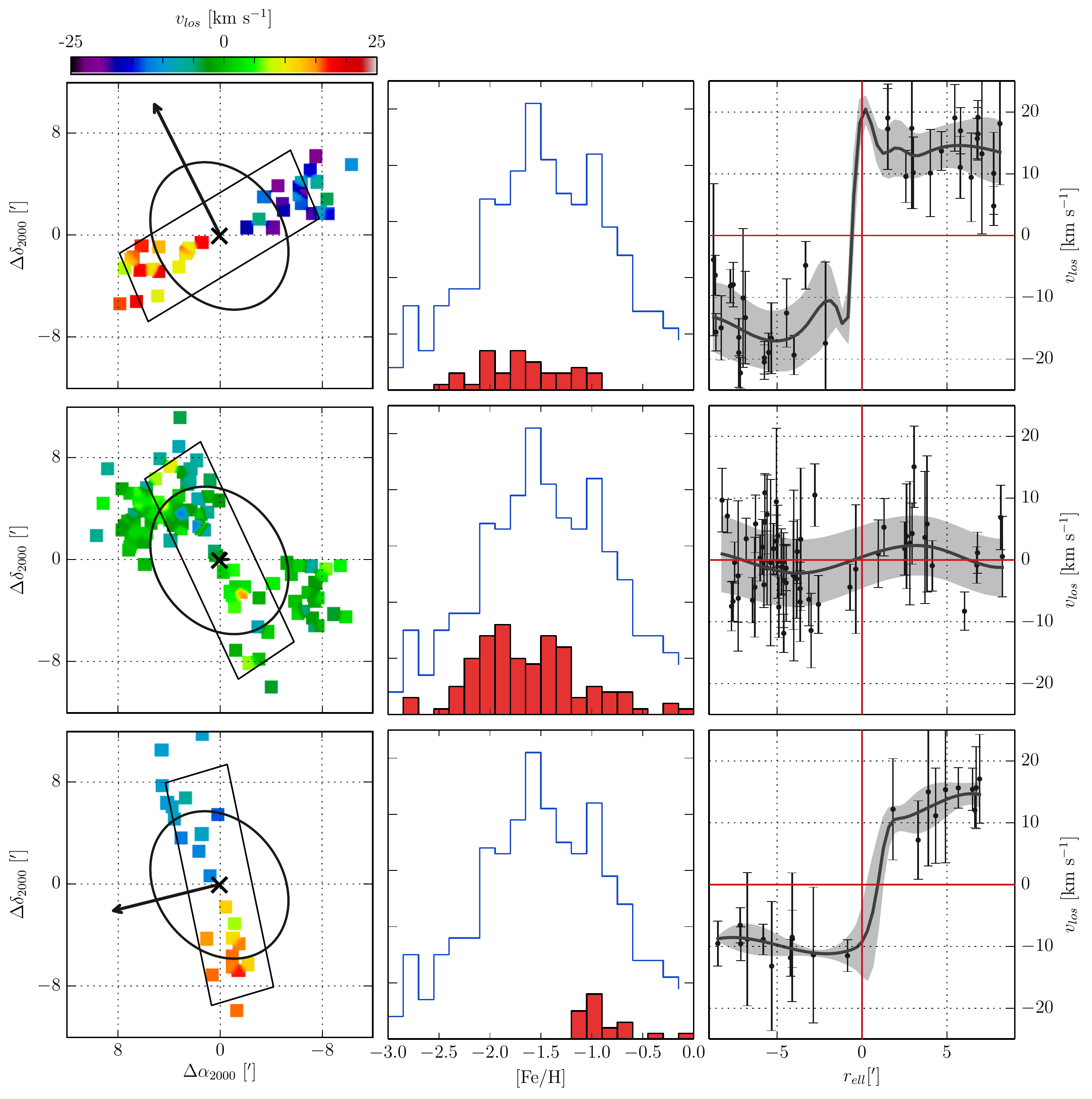}
\caption[Stars used for the chemo-kinematic analysis]{Same as Figure~\ref{fig:rawmaps}, for the different BSSs found.
The rows of panels from top to bottom correspond to streams BSS1, BSS2 and BSS3. Left panels show the \vlos~maps
with the black arrow representing the angular momentum $\mathbf{L}$. Middle panels show the metallicity distribution of
the stars shown in the left panels. Right panels plot the velocity of the subsample of stars along the stripe shown in
the left panels. {Velocity peaks present in the inner regions of the rotation curve of BSS1 are presumably produced by numerical
oscillations of the cubic spine fitted to the data.} Results from top to bottom are ordered by the average metallicity of the groups from the lower to
higher metallicity. The notation and line types are the same as in Figure~\ref{fig:rawmaps}.}
\label{fig:v_populations}
\end{center}
\end{figure*}

After applying \beacon~to the \citet{Ho2012} catalogue the systemic velocity of And II was found to be $v^{\rm
CM}_{los} = 192.2 \pm 0.6$ km s$^{-1}$, in good agreement with previous results from \citet{Kaliari2010} and
\citet{Ho2012}. This estimate is based on the 198 stars which where catalogued in three possible BSSs. The 333 remaining stars were found not to belong to any particular motion stream, but to follow the general prolate rotation pattern of the galaxy. The results are
shown in Figure~\ref{fig:v_populations}. We found three main chemo-kinematic components in And II. The first one (BSS1)
consists of 48 metal poor stars ($\langle \rm [Fe/H] \rangle = -2.0 \pm 0.5$) rotating roughly around the optical major
axis of the galaxy. Its kinetic PA, PA$_{\rm kin}=63^\circ \pm 3^\circ$, is fully consistent with the
photometric PA of the old population. The second one (BSS2) is a group with $\mathbf{L}$ nearly equal to zero, with
124 stars. The fact that these stars are distributed along the major axis of the galaxy, which is also the main rotation
axis of the galaxy, makes us think that this group is compatible with the first one. The component along the
line of sight of their velocities when they are close in projection to the rotation axis should be close to zero. In
addition, the metallicity distribution of both BSSs is similar. The last of the BSSs (BSS3) is composed of 26
more metal rich stars ($\langle \rm [Fe/H] \rangle = -1.0 \pm 0.4$) and shows a strong velocity gradient, but this
time along the north-south direction (PA$_{\rm kin} = 18^\circ \pm 2^\circ$). The general rotation pattern of the
galaxy could be therefore expressed as a sum of these two main components: a metal-poor population rotating
around the projected optical major axis of the galaxy and a more metal-rich one rotating roughly around the minor
axis. It is interesting to note that \beacon~found different rotation centres for each BSS. The centre of BSS1 was
consistent within errors with the optical centre found for the old population, while BSS3 was found to be rotating
around the optical centre of the intermediate-age stellar population (see Table~\ref{tab:Ellipses_promedio}).

For each BSS we set a stripe along the direction of maximum velocity gradient with the width of 3 arcmin. The \vlos~of
the stars within the stripe are shown as a function of the galactocentric elliptical radius in the right panels of
Figure~\ref{fig:v_populations}. Both populations rotate with similar velocities, although the rotation curve of BSS1
appears to be flatter than the one of BSS3. Using stars located at $r_h-0.5^\prime < r_{ell} < r_h+0.5^\prime$ along
the stripe, we measured an average \vlos$(r_h) = 16 \pm 3$ km s$^{-1}$ for BSS1 and \vlos$(r_h) = 13 \pm 4$ km s$^{-1}$
for BSS3. {The two populations were also found to have similar velocity gradients, only marginally larger for BSS3,
$2.24 \pm 0.22$ km s$^{-1}$ arcmin$^{-1}$, than for BSS1, $2.06 \pm 0.22$ km s$^{-1}$ arcmin$^{-1}$.} This was
determined by a linear regression of the stellar velocities, \vlos, as a function of galactocentric elliptical
distances, $r_{ell}$, along the stripe. {More specifically, a straight line was fitted to the data along the stripe
in the \vlos--$r_{ell}$ plane, obtaining its slope and y-intercept. The interception of these lines with the
\vlos~axis} gives an idea about possible offsets in the \vlos~of each BSS with respect to the ${v}^{\rm CM}_{los}$. The
best fit for the BSS1 gave a value ${v}^{zero, \rm BSS1}_{los} = 0.4 \pm 1.4$ km s$^{-1}$, which is compatible with the
systemic velocity of And II. Results for BSS3 were more intriguing. With ${v}^{zero, \rm BSS3}_{los} = 1.7 \pm 1.1$,
the centre of rotation of BSS3 appears to have a slightly higher \vlos.

\begin{figure}
\begin{center}
\includegraphics[ scale=0.7]{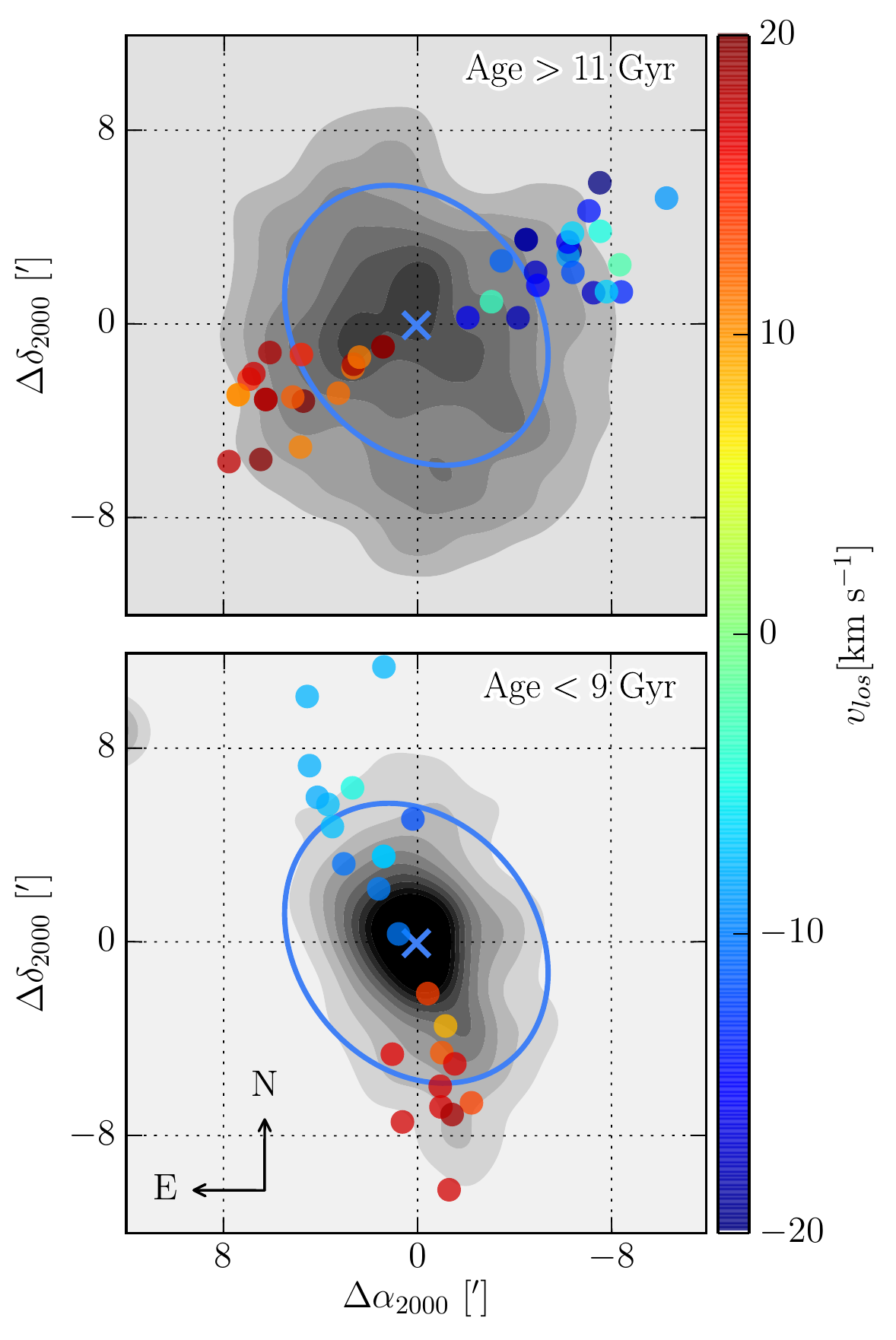}
\caption[]{Old and intermediate-age BSSs and surface density maps. Contour levels indicate the concentration of stars in
steps of 0.5$\sigma$ up to 4$\sigma$ in units of standard deviation, and are common for the two panels. Symbols and
line types coincide with those used in Figure~\ref{fig:Data}.}
\label{fig:Phot_BEACON}
\end{center}
\end{figure}

The results obtained through \beacon~were crossmatched with the sky positions of those from the photometry in order to
derive an approximate average age of the BSSs. We found 12 common stars between the BSS1 and our cleaned photometric
list, and only 10 for the BSS3. Using these stars we obtained an average age of $12.2 \pm 0.3$ Gyr for the BSS1, which
means that all its stars are older than 11 Gyr. On the other hand, BSS3 shows an average age of $8 \pm 2$ Gyr. One of
the BSS3 stars for which we have ages from photometry was older than 10 Gyr, indicating some possible contamination
from old stars in this BSS. This old star shows the lowest metallicity in BSS3 ($\rm [Fe/H] \sim-1.2$). With this low
metallicity, it is likely that the star was indeed born during the first star formation burst of And II, $\sim 13$ Gyr
ago (see Figure~\ref{fig:Andromeda_II_model}). Taking into account that the metallicity distribution of BSS3 peaks at
higher metallicities, $\rm [Fe/H] \sim -1.0$, we expect the total contamination by old stars to be in fact below
10\%. This small contamination could explain the observed difference between the photometric PA of the intermediate-age
population and the PA$_{\rm kin}$ of BSS3, although this difference decreases with galactocentric radius (see
Figure~\ref{fig:Ellipses_evolution}). To check how strong the impact of the old star contaminants could be, we removed
all stars with metallicities lower than $\rm [Fe/H] = -1$ and repeated all measurements. The remaining 19 stars give a
PA$_{\rm kin} = 19 \pm 3$), \vlos$(r_h) = 14 \pm 3$ km s$^{-1}$, a velocity gradient of $2.12 \pm 0.13$ km s$^{-1}$
arcmin$^{-1}$ and ${v}^{zero, \rm BSS3}_{los} = 1.8 \pm 0.7$. These values are compatible with the original ones
obtained for BSS3. Assuming therefore that BSS1 traces the rotation of the old stars and BSS3 of the intermediate-age
ones, we have overplotted their stars on the stellar density contours of the two populations from the photometry.
Results can be seen in Figure~\ref{fig:Phot_BEACON}.

And II rotates around its optical major axis, but this rotation signal may be composed of two slightly different
streaming motions. The first one consists of old stars rotating approximately around the projected optical major
axis of the galaxy. These stars are more numerous than the younger ones (see Figure~\ref{fig:Andromeda_II_model}),
giving the galaxy its well known dominant prolate rotation signal. This is consistent with the rounder shape of
the galaxy when considering only its older component, especially given the shape of the ellipses at small radii, which
are elongated and oriented following the observed velocity gradient in BSS1. Intermediate-age stars, on the other hand,
appear to rotate roughly around the projected optical minor axis, following the elongated shape of their surface density
distribution seen in Figure~\ref{fig:Density_Maps}. The centres of rotation of each population coincide with their
optical centres.

In addition to those described above, {trying other different parameters with \beacon~we} detected other
kinematically distinct components in And II {that are not shown in the figures}. The most remarkable one is a
group of metal-rich stars ($\rm [Fe/H]> -0.5$) located at the very centre of the galaxy, extending not further than
$\sim 2^\prime$ in galactocentric radius. We believe that these stars are the same as the ones causing the central
stellar overdensity. Their kinematics does not show any special features, with almost no velocity gradient and
\vlos$\sim 0$ km s$^{-1}$.

{This is not the first time that different kinematics are found for different stellar populations in dSphs. For
example, \citet{Tolstoy2004} found that splitting a sample of 308 RGB stars in the
Sculptor dSph Galaxy into metal-poor and metal-rich resulted in finding different $\sigma$(\vlos) for both subsamples:
$11\pm1$ and $7\pm1$ km s$^{-1}$. Even more convincing were the results for the Fornax dSph galaxy, where
\citet{Amorisco2012} found different velocity gradients of different stellar populations through Bayesian techniques.}

\subsection{Voronoi tessellation over resolved stellar populations}\label{Cap:Voronoi}

Two-dimensional maps of spectroscopically observed stars are very useful to look for any possible anisotropy. However,
creating such maps poses two main problems. The first one is the scarce number of stars typically available in
spectroscopic samples of dSphs. The second one is that the spatial coverage of these is usually not uniform. These two
circumstances make it impossible to keep a constant number of stars per unit of resolution or `pixel' in the maps. To
overcome this, we decided to apply Voronoi tessellation technique to the whole spectroscopic sample of \citet{Ho2012}.
This procedure creates cells with constant number of stars based on their position and surface density.

\begin{figure}
\begin{center}
\includegraphics[ scale=0.7]{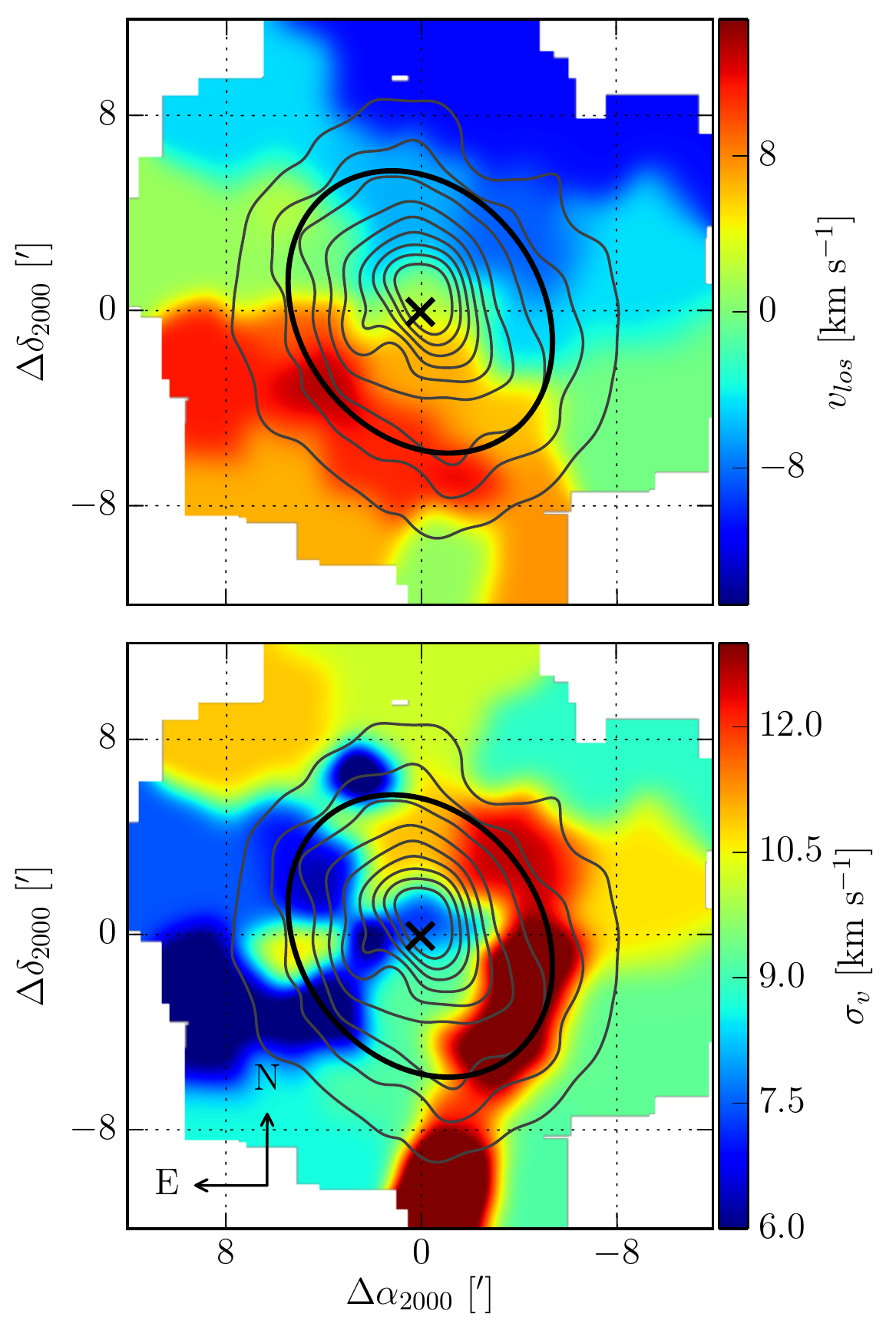}
\caption[Voronoi]{The map of \vlos~ (upper panel) and the velocity dispersion, $\sigma_v$ (lower panel) for the 544
stars from the \citet{Ho2012} sample. Contours represent stellar density from the total photometric sample (24562
stars). Symbols and line types coincide with those used in Figure~\ref{fig:Data}.}
\label{fig:voronoi}
\end{center}
\end{figure}

The Voronoi tessellation was performed using a modified version of the code provided by
\citet{Capellari_Voronoi}, adapted to work with resolved stars. We computed the velocity, \vlos, and the velocity
dispersion, $\sigma_v$, of the stars in the Voronoi cells. Several configurations were used, ranging from 5 to 25 stars
per cell. All experiments provided similar and fully consistent results. The configuration with 15 stars per cell,
smoothed with a Gaussian filter, is shown in Figure~\ref{fig:voronoi}. Interestingly, we detect a region of higher
velocity dispersion extending in the north-south direction. This high dispersion region shows twice larger dispersion
than measured in other regions of the galaxy ($\sigma_v \sim 14$ km s$^{-1}$) and overlaps with the elongation of the
galaxy towards the south direction. This indicates the presence of dynamically decoupled stars in that zone. Its curved
shape around the centre of the galaxy also suggests that, whereas these stars are not following the general rotation
pattern of the galaxy, they could be interacting {gravitationally with the galaxy}.

\section{Discussion}\label{Cap:Discussion}

The existence of two distinct stellar populations in And II, different in age and spatial distribution, has been
demonstrated by several works \citep{DaCosta2000, McConnachie2007, Weisz2014}. This feature could arise naturally from
a quiet secular evolution of the galaxy as a consequence of the star formation feedback and SN feedback. A first burst of star
formation would heat, or even expel, part of the gas in the galaxy. Once the gas is reaccreted and cooled in
the central parts of the galaxy, a second burst occurs.

{Besides internal processes such as feedback, other global or cosmic environmental factors could have interfered
with the SFH of And II. For example, the heating from the ultraviolet radiation from cosmic reionization could be also
responsible for the star formation quenching that occurred after the first burst. Evidence from quasar spectra indicates
that the Universe was fully reionized by $z \sim $ 6, corresponding to a look-back time of $\sim$12.7 Gyr
\citep{Becker2001}. This coincides approximately with the moment when the star formation starts to decrease, to be
completely stopped around 10.5 Gyr ago (Figure~\ref{fig:Andromeda_II_model}). The investigation of the individual
contribution of these two effects (reionization and feedback) to the SFH of And II would require complex simulations
for which initial conditions are still unknown. Yet, it is remarkable that after $\sim 1.5$ Gyr of inactivity, And II
started forming stars again. This would indicate that, in the case of secular evolution, the galaxy retained some gas
bound to its potential well, and that none of these processes could effectively suppress star formation in the
galaxy. However, none of these effects explains the prolate rotation of And II, neither why both populations show
slightly different kinematics.}

Another possibility would be that And II had recently interacted with M31 and that tidal forces during such interaction
would have caused its odd kinematics \citep{Ho2012}. However, this scenario seems improbable from the theoretical point
of view. $N$-body simulations show that tidal forces remove any strong rotation signal in dwarfs. Only very specific
inclinations of the orbit of And II around M31 could induce some rotation around the major or intermediate axis and
always smaller than the remnant rotation around the minor axis. The large amplitude of the rotation signal compared to
the velocity dispersion was also not reproduced in the case of such tidal stirring scenarios
\citep{Lokas2014b, Ebrova2015}.

To overcome these inconsistencies, \citet{Lokas2014b} proposed a scenario, later developed by \citet{Ebrova2015} and
\citet{Fouquet2017}, in which And II is a result of a merger between two late-type galaxies of similar mass. In this
scenario, the prolate rotation results naturally from the conservation of the angular momenta of the progenitors along
the direction of the merger. Previous observational evidence seems to support this scenario. The picture it proposes is
consistent with SFH of And II \citep[][Hidalgo et al. in preparation]{Weisz2014}, the truncated radial profile of the old
population \citep{McConnachie2007}, or the lower dispersion regions identified by \citet{Amorisco2014}. On the other
hand, the lack of statistical differences between the metal-rich and metal-poor populations defined by
\citet{Ho2012} would suggest that both stellar populations are not kinematically different. If And II is a result of
a major merger, also kinematic signatures are expected to remain in the galaxy.

In the present paper, we have tried to put together all the pieces of information that we have in order to conclude
whether there exists, or not, strong enough evidence of past interaction of And II with another system. For example,
we have tried to reproduce the stellar stream advocated by \citet{Amorisco2014}. By giving more statistical weight to
the radial coordinate, $r_{ell}$, during the clustering process in \beacon, we can obtain BSSs with elliptical or
circular shape, similar in their spatial distribution to the alleged stream. However, we could not conclude that these
stars were in fact a part of a stream. Our elliptical BSSs were in reality just areas with a lower velocity dispersion.
The metallicity distribution of their stars was very similar to the global metallicity of And II and the \vlos~of their
stars followed closely the average \vlos~of the whole sample of stars at these galactocentric radii, i.e. the prolate
rotation signal. We think that nothing clear in
favour or against the merger scenario can be concluded from these results.

On the other hand, the results obtained here differ from those found by \citet{Ho2012}, in spite of the fact that we
use the same data set. In that work, the authors divide the stars into a metal-poor ([Fe/H]$ < -1.39$) and metal-rich
([Fe/H]$ > -1.39$) subsamples, and conclude that no statistical difference can be found between these two groups of
stars. Inspecting the SFH of the galaxy (Figure~\ref{fig:Andromeda_II_model}), we find that a higher metallicity
cut ([Fe/H]$ \sim -1.1$) should be taken in order to separate both populations more accurately. In order to quantify the
contamination by old stars in the metal-rich group selected by \citet{Ho2012}, we crossmatched the whole spectroscopic
list with the \citet{McConnachie2007} photometric sample with assigned ages. We found 192 common stars from which 121
had a photometric assigned age older than 10 Gyr, resulting in an average age of $10.2\pm 1.9$ Gyr. When using the
clean photometric list only 36 stars were matched, with an average age of $9 \pm 2$ Gyr. In this case, 10 stars had
ages older than 10 Gyr in their photometric counterpart. We think that this contamination by old stars could be
hiding the intrinsic kinematic features of the intermediate-age population in the metal-rich population of
\citet{Ho2012}.

Our results indicate that And II is not {fully relaxed} and that it possesses at least two centres of rotation. If we
identify BSS1 and BSS3 with the old and the intermediate-age stellar populations respectively, it appears that
old stars rotate around the projected optical major axis of the system, while the younger ones do it around the
projected optical minor axis. The two populations do not share optical centres, and their kinematic centres also differ
in their \vlos. The velocity dispersion is not constant nor symmetric along the body of the galaxy. It is interesting
to note that the region with higher $\sigma_v \sim 14$ km s$^{-1}$ extends in the direction of the higher elongation of
the galaxy body at the large radii, overlapping partly with the intermediate-age population. Stars in this region
belonging to BSS3 show higher value of \vlos~than their counterparts at the other side of the CM, which may
indicate some residual odd kinematics. The radial density profiles are also bizarre: the old population is truncated at
large radii ($1650\pm40$), while the younger one seems to be perturbed, showing a large elongated tail towards
the south-west direction.

All these features could have arisen from interactions with other systems. We believe that a major merger that took
place $\sim$ 9-10 Gyr ago is the simplest way to explain all these odd observational properties. The two involved
galaxies would have paused their star formation due to the feedback during and after the first star formation burst,
$\sim$ 12 Gyr ago. This, together with tidal forces during first close passages, could have stripped and expelled the
gas from both galaxies. Once the fusion was completed around $8.5$ Gyr ago, the strengthened potential well would have
recaptured the gas in the centre of the galaxy. This would have started the star formation again. This scenario could
also explain the differences between the kinematics of the old and intermediate-age stars.

A similar scenario was tested with $N$-body/hydrodynamical simulations including gas dynamics and star formation by
\citet{Fouquet2017}. In these simulations the two galaxies were quiescent at the moment of the collision, and it was
the compression of the gas in the centre after the merger that produced the second burst of star formation. However, in
these simulations a large amount of gas remains in the galaxy after the merger. The authors postulated that a recent
passage of And II by M31 (4.8 Gyr ago) would have blown away the remaining gas from And II via tidal and ram
pressure stripping. This would have stopped the star formation and produce the gas-deficient galaxy we observe today
\citep{Grcevich2009}. Although this scenario is more difficult to test, a close passage could be responsible for
asymmetric tidal perturbations in the kinematics of the stars that can last up to several Gyr. The observed
asymmetric distribution of $\sigma_v$ (see Figure~\ref{fig:voronoi}) could be therefore caused by this kind of
interaction. Furthermore, And II could have suffered from more than one interaction with other systems. It is also
possible that it recaptured gas with different angular momentum previously expelled from the galaxy due to
supernovae feedback, causing the observed kinematic perturbations. All these possible scenarios call for further
development of
the model of \citet{Fouquet2017} including improvements in the treatment of supernovae and star formation feedback.

\section{Summary and conclusions}
\label{Cap:Summary}

We have analyzed And II to unprecedented detail by combining photometry and spectroscopy of its stars and using
\beacon. We have obtained the ages, metallicities and kinematic properties of its two stellar populations. These
populations differ in all the properties we have measured in the present work, namely ages, metallicities, spatial
distribution and kinematics. Here we list the main results of this work.

\begin{itemize}
 \item The first stellar population consists of old ($\gtrsim 11$ Gyr), relatively metal-poor ([Fe/H]$< -1.25$) stars.

 \item The second stellar population is composed of younger ($\lesssim 9$ Gyr) and more metal-rich ([Fe/H]$> -1.1$)
stars. This second population could actually form in two star formation bursts that occurred $\sim 8$ Gyr and $\sim
6.25$ Gyr ago.

 \item The old population is distributed uniformly and with constant density following a round distribution up to
large radii ($r_{\rm c} = 8.7^\prime\pm0.2^ \prime$,  $1650 \pm40$ pc). The radial density profile is better fitted by
truncated profiles (with S\'ersic index, $n = 0.43\pm0.03$).

 \item The intermediate-age population is more concentrated ($r_{\rm c} = 1.3^\prime\pm0.5^\prime$, $240\pm90$ pc)
and forms a stellar overdensity observed in the central parts of the galaxy. Its stars form a large
structure elongated along the optical major axis of the galaxy, and then the north-south direction.

 \item  Isopleth contours of the stellar surface density show significant variations as a function of the galactocentric
distance. This may indicate that the galaxy is not {fully relaxed}, or that it has undergone some tidal
interaction with another galaxy.

 \item And II appears to have at least two different dynamical components. Stars of these components have been
identified with the old and the intermediate-age populations.

 \item Kinematics of both stellar populations differs: they do not share the same centre of rotation or kinetic PA.

 \item Old stars appear to rotate around the projected optical major axis of the galaxy, with \vlos$(r_h) = 16 \pm 3$
km s$^{-1}$ and a velocity gradient of $2.06 \pm 0.21$ km s$^{-1}$ arcmin$^{-1}$.

 \item Intermediate-age stars appear to rotate roughly around the projected optical minor axis of the galaxy, with a
velocity gradient of $2.24 \pm 0.22$ km s$^{-1}$ arcmin$^{-1}$.

 \item The velocity dispersion is not constant nor symmetric along the body of And II, showing a maximum of $\sigma_v
\sim 14$ km s$^{-1}$ located in the south-west parts of the galaxy. This may indicate the presence of dynamically
decoupled stars in this region.

\end{itemize}

All these results, together with the bimodal SFH showing a clear quenching of star formation between $\sim$10.5 and
$\sim$9.5 Gyr ago, support a scenario in which And II is the result of a past merger that occurred at redshift $z\sim
1.75$ ($\sim 10$ Gyr ago).

\section*{Acknowledgements}

We are grateful to Dr. D. Weisz, Dr. A. McConnachie and Dr. N. Ho and collaborators for generously providing their data for this
project. {The authors thank the anonymous referee for the comments that have helped to improve this paper.} AdP also thanks S. Bertran de Lis for her support and help during this project. This research was supported by
the Polish National Science Centre under grant 2013/10/A/ST9/00023.

%%%%%%%%%%%%%%%%%%%%%%%%%%%%%%%%%%%%%%%%%%%%%%%%%%



% \bibliographystyle{mnras}
% \bibliography{Stellar_distributions_Andromedaii_3} % if your bibtex file is called example.bib


%%%%%%%%%%%%%%%%%%%%%%%%%%%%%%%%%%%%%%%%%%%%%%%%%%

\bsp	% typesetting comment
\label{lastpage}
\end{document}